\documentclass[11pt]{article}
\usepackage{amsmath, amssymb}
\usepackage{slashed}
\usepackage{verbatim}
\usepackage{latexsym}
\usepackage{euscript}
\usepackage{amsfonts}
\usepackage{verbatim}
\usepackage{cancel}
\usepackage{soul}
\usepackage[normalem]{ulem}
\usepackage[]{array}
\usepackage[]{mathrsfs}
\usepackage[utf8]{inputenc}
\usepackage[T1]{fontenc}
\usepackage{makeidx}
\makeindex
\usepackage{graphicx}
\usepackage{amsmath}
\usepackage{amssymb}
\usepackage{amsxtra}
\usepackage{color}
\def\be{\begin{eqnarray}}
\def\ee{\end{eqnarray}}
\def\0{\nonumber}

\def\tr{{\rm tr}}

\def\sfg{{\sf g}}

\usepackage{slashed}
\usepackage{verbatim}
\usepackage{latexsym}\usepackage{color}
\usepackage{euscript}
\usepackage[curve]{xypic}
\newcommand\EW{\EuScript{W}}

\newcommand\ET{\EuScript{T}}

\newcommand\ED{\EuScript{D}}

\def\sfD{{\sf D}}

\def\sfd{{\sf d}}\def\sfg{{\sf g}}

\normalfont\large
\begin{document}
\vskip 1cm
\begin{flushright}
{SISSA/13/2026/FISI}
\end{flushright}
\vskip 1cm
\begin{center}

{\LARGE Conformal symmetry, SM and Gravity }

\vskip 1cm

{\large  L.~Bonora$^{a}$\footnote{email:bonora@sissa.it},\\
\textit{${}^{a}$ International School for Advanced Studies (SISSA),\\Via
Bonomea 265, 34136 Trieste, Italy}
}

\end{center}
\vskip2cm
{
{\bf Abstract}. This paper is a bottom up attempt to incorporate the standard model and general relativity in a unique quantum field theory. The tentative model presented here in particular is free of chiral gauge and gravitational anomalies that appear in the divergence of currents, and in the divergence and trace of the energy-momentum tensor when the SM matter couples to gravity. The fermion spectrum is composed of two multiplets, the SM (left) multiplet and a mirror copy (right) with opposite handedness. The right  multiplet is interpreted as describing the dark matter world. The natural symmetry of the theory is enlarged to incorporate also Weyl invariance, by introducing one or more dilaton fields. After the cosmological and theoretical motivations, the necessary formalism is introduced for algebraic renormalization: gauge fixings, ghosts, propagators and vertices and their interplay in guaranteeing the conditions for convergence of the subtracted amplitudes according to the BPHZL scheme, the Slavnov-Taylor identity and the relevant enlarged BRST symmetry. The corresponding (conformal) cohomology is analyzed and found to be trivial: there are no non-trivial even trace anomalies in theories with dilatons, but there are plenty of trivial ones, which require corresponding counterterms in the effective action. It is shown that such counterterms can play an important role in freeing the theory of unphysical particles.

\section{Introduction}

The purpose of this paper is to analyse  the consequences of incorporating in the same quantm field theory  the Standard Model (SM) of particle physics and General Relativity (GR), the latter considered as the classical version of a field theory to be quantized. The approach is bottom-up, that is the resulting theory does not descend from a fundamental one, via for instance compactification of extra dimensions, like in superstring theories, but is based on the idea of using all possible information one can gather when putting together SM and GR and going to the regime where both are quantum theories. 
In \cite{BG24} an explicit model of this type was proposed. It was shown there that some basic information comes from the anomaly analysis. This is due to the fact that the SM is formulated in terms of Weyl fermions, and Weyl fermions are a natural source of chiral anomalies. Anomalies belong to two large classes, obstructive  (type O) anomalies and non-obstructive  (type NO) anomalies. The former are dangerous as they correspond to the non-existence of the propagators for Weyl fermions, therefore they constitute a hazard for field theory quantization. The latter do not lead to such negative consequences and are simply honest quantum effects. The SM is so constructed that all type O anomalies cancel out. This is not completely true when we couple the SM Weyl fermions to a metric as in the present approach. Chiral gravitational anomalies vanish. There are mixed gauge-gravity anomalies, whose coefficients however sum up to zero. There remain odd parity trace anomalies due to the metric and the gauge fields. Most of them cancel, but the trace anomaly due to the $SU(2)$ gauge field does not. This anomaly is type O, it signals the lack of a propagator for the Weyl fermions of the theory, as stated by the family's index theorem of Atiyah and Singer, \cite{AS}. The conclusion of the first part of \cite{BG24} is that when we incorporate GR and SM in a unique teory we meet this first obstacle, which must be overcome if we want to arrive at a theory in which quantization makes sense.

The second part of \cite{BG24} proposes a solution for the previous problem: a left-right symmetric model made of two copies of the SM which mirror each other, that is fermions representing left-handed particles and right-handed antiparticles in one side are mirrored by fermions representing right-handed particles and left-handed antiparticles on the other side, and viceversa. This arrangement cancels all O type anomalies. The left side is just a copy of the SM, coupled to an $SU(3)_L \times SU(2)\times U(1)_L$. The right side is another copy of the SM, but with opposite chiralities, and coupled to an $SU(3)_R\times SU(2)\times U(1)_R$ gauge group. In other words the two halves have each its own strong and $U(1)$ interactions while the weak forces are in common. In \cite{BG24} the model was characterized by two different metrics (left and right). Subsequently , in \cite{bonora25} the model was simplified by considering only one metric, and this simplification is assumed also in the present paper. Therefore in the resulting model, denoted $\cal T$, whose action is presented in the next section,  the two halves share the same weak and gravitational forces. In \cite{bonora25} a closer analysis of $\cal T$ was initiated, but still mostly in the form of a research design. One idea put forward there is that if the left sector represents the standard model, the most natural interpretation of the 
right sector is that it represents the dark matter world. The second idea arises from the observation that the fermion and gauge sectors of the SM, as well as the Yukawa couplings, are conformal invariant. It is natural to imagine that a more fundamental theory, from which $\cal T$ can descend, is a conformal invariant one. It is easy to transform any ordinary theory which is not conformal invariant into a conformal invariant one with the help of a dilaton field (two, one left and one right, in the case of $\cal T$). This led in \cite{BG24} to $\cal TW$, the conformal version of $\cal T$.  

A conformal field theory of matter and gravity may solve two big problems. The first problem arises from the idea itself of juxtaposing two field theories like SM and GR. Any quantum field theory is characterized by its own vacuum energy. The trouble is that SM and gravity have hugely different vacuum energies, so that it simply does not make sense to quantize them over the same vacuum. As was pointed out in \cite{bonora25} this dramatic issue can be solved by the presence of a dilaton. The second problem is that of unitarity and renormalizability of $\cal TW$. The main idea was briefly outlined in \cite{bonora25} and will be to a large extent (although not completely) developed in this paper. The main steps will consist in defining for $\cal TW$ the framework for algebraic renormaliztion within the BPHZL regularization scheme: setting the formalism, fixing the gauges of diffeomorphisms and Weyl transformations, computing the propagators and vertices, verifying the Lowenstein conditions for convergence of the 1PI subtracted amplitudes, defining the Slavnov-Taylor (ST) identities, the gauge fixing and ghost equations. This will allow us to define the BRST enlarged operator and the relevant cohomology analysis, which is the preliminary step for the renormalization process. We shall analyze in depth this cohomology. It turns out to be trivial: no true anomalies can be found, only trivial anomalies or coboundaries are allowed. The coboundaries representing trivial trace anomalies are however particularly important. It will be shown how the counterterms needed to cancel them may be used to exclude the presence of unphysical particles.

The paper is organized as follows. To render it as self-contained as possible, in section 2 we review in detail the model put forward in \cite{BG24,bonora25}, and, in section 3, the interpretation of the mirror sector as describing the dark matter world. 
In section 4 the Weyl geometry is introduced and applied to the $\cal T$ theory, together with the motivation for viewing this symmetry as a fundamental one. Of course, as made explicit in the motivations, all this make sense if we can quantize this theory and the quantization makes sense, that is guarantees renormalizability and unitarity. In section 5 the problems underlying the quantization of a gravitational theory are outlined (dropping for simplicity matter fields) and all the indispensable formalism is introduced for algebraic renormalization \cite{piguetsorella}: gauge fixings, ghosts, propagators and vertices and their interplay in guaranteeing the conditions for convergence of the subtracted amplitudes according to the BPHZL scheme. It is shown that, at least in a particular gauge these conditions (proposed by Lowenstein) are satisfied. Then the Slavnov-Taylor (ST) identity is introduced together with the enlarged BRST transformation (the linearized ST) and its gauge independence. In section 6 the cohomology of the enlarged BRST operator is analyzed and completely solved: all invariants are identified and it is shown that no non-trivial conformal trace anomalies are present in the theory. It is shown however that trivial ones (coboundaries) are nevertheless very important and may be relevant in relation to the problem of unitarity,for they may cancel the quartic derivative terms for special values of the dilaton(s), as shown in section 7. In section 8 the trace anomaly analysis is repeated within the ordinary, down-to-earth, formalism (as opposed to the algebraic renormalization one) and confirm the results of the previous section. Finally section 9 is devoted to some (temporary) conclusions.

\section{A left-right symmetric model} 

This first section is devoted to a concise description of a left-right symmetric model of matter minimally coupled to gravity.
It was introduced in \cite{BG24} with two metrics. Here a simplified version is considered, in which there is only one metric. The fermion matter part is based on the same multiplet as the MSM with the addition of a right-handed sterile neutrino. In the usual SM notation it is 
\be
\begin{matrix} {\sf G}/fields & \quad SU(3)\quad &\quad SU(2)\quad &\quad U(1)\quad\\
\left( \begin{matrix} u \\ d\end{matrix} \right)_{\! L} & 3&2&\frac 16\\
{(u_R)^c} &  \bar 3 &1&-\frac 23\\
{(d_R)^c} & \bar 3 &1&\frac 13\\
\left( \begin{matrix} \nu_e \\ e\end{matrix} \right)_{\! L} & 1&2&-\frac 12\\
{(e_R)^c} &  1 &1&1\\
{(\nu_R)^c} &  1 &1&0
\end{matrix}\label{Lspectrum}
\ee
where $ X^c$ represents the Lorentz conjugate spinor of $X$, i.e. $X^c=\gamma_0 C X^\ast$. This multiplet couples  to the $SU(3)_L\times SU(2)\times U(1)_L$ gauge fields.as well as to a gravitational metric and connection. The main criterion for the definition of the model is the absence of type-O(obstructive) anomalies. In \cite{BG24} it was shown that all such anomalies cancel out except for  4 units of the trace anomaly with density $F\ast F$, due to the gauge field   $F\equiv F^{\mathsf su(2)}$,  computed in the doublet representation of $\mathsf su(2)$.

In the multiplet \eqref{Lspectrum} the doublet $\left(\begin{matrix} u\\ d\end{matrix}\right)_L$  describes left-handed particles and right-handed antiparticles, while the singlets $u_R$ and $d_R$ represent right-handed particles and left-handed antiparticles, and similarly for the leptons.

The main difference with the MSM is that the spectrum is completed by a right-handed multiplet
\be
\begin{matrix} {\sf G}/fields & \quad SU(3)\quad &\quad SU(2)\quad &\quad U(1)\quad\\
\left( \begin{matrix} u' \\ d'\end{matrix} \right)_{\! R} & 3&2&\frac 16\\
{(u'_L)^c} &  \bar 3 &1&-\frac 23\\
{(d'_L)^c} & \bar 3 &1&\frac 13\\
\left( \begin{matrix} \nu'_e \\ e'\end{matrix} \right)_{\! R} & 1&2&-\frac 12\\
{(e'_L)^c} &  1 &1&1\\
{(\nu'_L)^c} &  1 &1&0
\end{matrix}\label{Rspectrum}
\ee
coupled to the gravitational metric and connection. It also couples to the $SU(3)_R\times SU(2)\times U(1)_R$ gauge fields. The anomaly analysis of this mirror multiplet\footnote{A mirror sector of the SM has been considered earlier in the literature, see \cite{hodges93,mohapatra97,bai14}, with various purposes related to cosmology.} is the same as for the left-handed one except for the sign of the trace anomaly due to the gauge field   $F\equiv F^{\mathsf su(2)}$,  which is opposite. Therefore the overall sum of the anomalies of the system vanishes.

In the multiplet \eqref{Rspectrum} the doublet $\left(\begin{matrix} u'\\ d'\end{matrix}\right)_R$  describes right-handed particles and left-handed antiparticles, while the singlets $u'_L$ and $d'_L$ represent left-handed particles and right-handed antiparticles, and similarly for the leptons.

Of course we should consider three families of left-handed and three families of right-handed fermions. But since the physics that intertwines different families will not be discussed here, one single family will do.

We shall call these two intertwined theories, with field content \eqref{Lspectrum} and \eqref{Rspectrum}, ${\cal T}_L$ and ${\cal T}_R$, respectively. The main features to be called to the reader's attention is that both multiplets interact with te same metric and the same $SU(2)$ gauge fields, but with different $SU(3)\times U(1)$ gauge fields on the left and the right. This is what makes the overall theory free of type O anomalies. We denote it simply by ${\cal T}={\cal T}_L \cup {\cal T}_R$. The symbol $\cup$ is to recall that the two half theories have the metric and the SU(2) gauge potentials in common.

{\bf Important.} Both multiplets couple to the same $SU(2)$  gauge fields. Only in this case do all obstructive anomalies cancel! We remark that, since, contrary to \cite{BG24} there is only one metric, the presence of the sterile neutrinos
$\nu_R$ and $\nu'_L$ is not necessary in order to cancel all type O anomalies.
\vskip 0.3cm
Let us see now explicitly the various possible pieces of the relevant actions. We start from the fermion kinetic actions. On the right sector we have 
\be
S_f^{(+)}\equiv S_{fR}
&=&\int d^4 x \, \left(\sqrt{g}\, i\overline {{\psi'}_{R}}  
\gamma^a
 e_a^{\mu}
\left(\ED_\mu^{(+)}+\frac 12 \omega_\mu \right)\psi'_{R}\right)( x)\label{fermionaction+}
\ee}
where $\psi'_R$ represents the right-handed multiplet \eqref{Rspectrum}, and 
\be
\ED^{(+)}_\mu=\partial_\mu -i{\sfg}_X^+ X^{(+)}_\mu -i{\sfg}_WW_\mu -i{\sfg}_B^+B_\mu^{(+)} \label{XWQ+}
\ee
According to the standard notation $ \omega_\mu=\omega_\mu^{ab}\Sigma_{ab}$ represents the spin connection corresponding to the metric $g$ and $\Sigma_{ab}$ the anti-hermitean Lorentz generators. For the left sector 
\be
S_f^{(-)}\equiv S_{fL}
&=&\int d^4 x \, \left(\sqrt{g}\, i\overline {{\psi}_L}  
\gamma^a
 e_a^{\mu}
\left(\ED^{(-)}_\mu+\frac 12 \omega_\mu\right)\psi_{L}\right)( x)\label{fermionaction-}
\ee
where $\psi_L$ represents the left-handed multiplet \eqref{Lspectrum}, and
\be
\ED^{(-)}_\mu=\partial_\mu -i{\sfg}_X^- X^{(-)}_\mu -i{\sfg}_WW_\mu -i{\sfg}_B^-B_\mu^{(-)} \label{XWQ-}
\ee
The symbols $X^{(\pm)}_\mu,W_\mu,B^{(\pm)}_\mu$ refer to the $SU(3)_{R/L}, SU(2)$ and $U(1)_{R/L}$ potentials, respectively, and take values in the corresponding Lie algebras, whose generators are taken to be Hermitean.  Each potential has its own distinct coupling to the fermions. The couplings here have  been made explicit through a redefinition of the potentials.

It is perhaps useful to recall that the symbol such as $(\psi_R)^c$ (for instance $(u_R)^c, (d_R)^c, ...$) can be rewritten as
\be
 (\psi_R)^c= \gamma^0 C \psi_R^*= \gamma^0 C P_R^* \psi^*=P_L \gamma^0 C\psi^*=P_L \psi^c = (\psi^c)_L.\label{psiLpsiR}
\ee
Inserted into the kinetic term, it gives
\be
\int d^4x\sqrt{g}\, \overline {(\psi^c)_L} \,\gamma^\mu(\partial_\mu +\frac 12 \omega_\mu )(\psi^c)_L=\int d^4x\sqrt{g}\, \overline {\psi_R}\, \gamma^\mu(\partial_\mu +\frac 12 \omega_\mu )\psi_R\label{LcR}
\ee
as can be seen by taking account that $\Sigma_{ab}$ are anti-hermitean, using an overall transposition and a partial integration. Therefore the kinetic term of the multiplet \eqref{Lspectrum}, coupled only to the metric, splits into 16 independent Weyl fermion kinetic terms, 8 left-handed and 8 right-handed, with opposite contribution to the odd parity trace anomaly. 

Next let us consider the gauge field sector. The SU(2) gauge field action has the usual form
\be
S^{\tiny{SU(2)}}_g =- \frac 1{4{\sf g}_W^2}  \int d^4 x \,\sqrt{g} \, \tr \left( g^{\mu\mu'} g ^{\nu\nu'}  F_{\mu\nu } F_{\mu'\nu'} \right)\label{actiongf+-}
\ee
where $F_{\mu\nu }=  \partial_\mu W_\nu- \partial_\nu W_\mu +i[ W_\mu, W_\nu]$ is the curvature of the SU(2) gauge field\footnote{In \cite{BG24} two $SU(2)$ gauge couplings were introduced, one for each sector; however the cancelation of $SU(2)$ gauge-induced odd trace anomalies requires that there be only one coupling.}.

For the groups $ SU(3)_L\times SU(3)_R$ and $ U(1)_L\times U(1)_R$ we have instead
$S_{g}^{(+)}+ S_{g}^{(-)}$ with
\be
S_{g}^{(\pm)} =- \frac 1{4 (\sfg^{\pm}_X)^2 }\int d^4 x \,\sqrt{g} \, \tr \left( g^{\mu\mu'} g^{\nu\nu'}  X^{(\pm)}_{\mu\nu } X^{(\pm)}_{\mu'\nu'} \right) - \frac 1{4 (\sfg^{\pm}_B)^2 }\int d^4 x \,\sqrt{g} \,  g^{\mu\mu'} g^{\nu\nu'}  B^{(\pm)}_{\mu\nu } B^{(\pm)}_{\mu'\nu'}  \label{actiongfae+-2}
\ee
where $ X^{(\pm)}_{\mu\nu }=  \partial_\mu X_\nu ^{(\pm)}-   \partial_\nu X_\mu ^{(\pm)}+i[ X_\mu^{(\pm)}, X_\nu^{(\pm)}]$  denotes the curvatures of a gauge field with values in the Lie algebras of $SU(3)_R$ and $SU(3)_L$, respectively, and $  B^{(\pm)}_{\mu\nu }=  \partial_\mu B^{(\pm)}_\nu-\partial_\nu B_\mu^{(\pm)}$ denotes the field strengths with values in the Lie algebras of $ U(1)_R$ and $U(1)_L$, respectively. The gauge couplings  can be absorbed, as usual, in a redefinition of the respective gauge potentials, as is done in \eqref{fermionaction+} and \eqref{fermionaction-}.

The action for the metric is the usual EH action with different cosmological constants  in the left and right sector 
\be
S^{(\pm)}_{EH}=- \frac 1{2\kappa}   \int d^4 x \,\sqrt{g} \left( R+ {\mathfrak c}_\pm\right)\label{EH+-}
\ee
where $R$ is the Ricci scalar, $\kappa$ the  gravitational constant and ${\mathfrak c}_\pm$ the left/right cosmological constant. 

In the MSM also a couple $H_\pm$ of complex scalar fields is needed. They are minimally coupled to the metric $ g_{\mu\nu}$ and form a doublet under $SU(2)$. The corresponding actions in the two sectors are
\be
S_{d}^{(\pm)}= \int d^4x \,\sqrt{g} \left[ g^{\mu\nu} \ED_\mu H_\pm^\dagger \ED_\nu H_\pm -  M_\pm^2 H_\pm^\dagger H_\pm -\frac {\lambda_\pm}4 \left(H_\pm^\dagger H_\pm\right)^2\right]\label{Hscalar+-}
\ee
where $\ED_\mu = \partial_\mu -i\sfg W_\mu$, and $W_\mu$ is the $SU(2)$ gauge field.

The above actions represent matter minimally coupled to the metric and to gauge potentials. Now we need action terms that represent the interaction among matter fields. They are provided by Yukawa couplings. They split into left and right parts. For instance, for SU(2) doublets we have
\be
S_{YdL} = \frac {y^-_{H_d}}2 \int d^4 x \,\sqrt{g}\left( \overline{\psi_{dL}}\,{H}_{d-} \chi_{sR}\right)+ h.c.\label{yukawadL+conj}
\ee
where $\psi_{dL}$ is a left-handed SU(2) doublet, $  H_{d-}$ is also an $SU(2)$ doublet, conjugate to the $\psi_{dL}$ one in the inner product of the $SU(2)$ doublet representation space, while $\chi_{sR}$ is a right-handed singlet, all of them belonging to ${\cal T}_L$. Similarly, for ${\cal T}_R$,
\be
S_{YdR} = \frac {y^+_{H_d}}2 \int d^4x \,\sqrt{g}\left( \overline{\psi'_{dR}}\,{H}_{d+} \chi'_{sL}\right)+ h.c. \label{yukawadR+conj}
\ee

Now we write $S_f= S_f^{(+)} +  S_f^{(-)} $, $S_g=S^{SU(2)}_g+ S_g^{(+)} +  S_g^{(-)} $, $S_{EH}= S_{EH}^{(+)} +  S_{EH}^{(-)} $, $S_d= S_d^{(+)} +  S_d^{(-)} $ and $S_Y= S_{YdL} +  S_{YdR} $. Then for the total action of  our model minimally coupled to gravity we can tentatively set
\be
S=S_f+S_g+S_{EH} +S_{d} + S_Y \label{totalaction}
\ee
Let us summarize the symmetries of this theory. It is invariant under $SU(2)$, as well as  $ SU(3)_L\times SU(3)_R$ and $ U(1)_L\times U(1)_R$, gauge transformations. It is also invariant under diffeomorphisms and local Lorentz transformations. As for discrete symmetries, each term of the sum \eqref{totalaction} is $CP$ and $T$ invariant in the left and right sector separately; but in general it is not $P$ and $C$ invariant. $P$ invariance of the overall $S$ would require (which will not be the case here) that all constants and masses appearing in $S$ with labels + and - be equal, i.e. ${\mathfrak c}_-= {\mathfrak c}_+$, etc. 

However left and right sectors separately have the same type of symmetries (the same symmetry only for diffeomorphisms and $SU(2)$ gauge). For this reason we may dub $\cal T$ a left-right or chirally symmetric theory (in a form, however, much looser than parity). For conciseness, we shall call left the fields of ${\cal T}_L$ and right the fields of ${\cal T}_R$, of course with the exclusion of the metric and the $SU(2)$ gauge fields.

As was mentioned above and shown in \cite{BG24}, both ${\cal T}_L$ and ${\cal T}_R$ are separately free of O-type anomalies, except for the trace anomaly whose density is $\sim F * F$. Putting together the two halves cancels also this anomaly.
Of ourse while the MSM is known to be a well-defined field theory, renormalizable and unitary, we do not have any similar claim for the theory presented here. It complies however with the first essential requirement for an effective theory: it is free of obstructive anomalies, so that all the propagators and vertices are well defined and a perturbative quantization can be carried out.  Concerning renormalization and unitarity, it is something we would like to broach in the rest of the paper, but, before, let us see a possible interpretation of the right sector.
 
\section{Dark matter?}

The theory $\cal T$ described by $S$ splits into two halves, each with distinct scalar and fermion matter components. Also the gauge groups $ SU(3)_L\times SU(3)_R$ and $ U(1)_L\times U(1)_R$, respectively, are distinct, while the metric and the $SU(2)$ gauge fields are the same on both sides. The left sector and the right sector interact with each other only via the latter fields and in no other way. For instance the left quarks (left-handed particles and right-handed antiparticles) $\left(\begin{matrix} u_L\\ d_L\end{matrix}\right)$   interact among themselves strongly via the $ SU(3)_L$ gauge bosons and electromagnetically via the $ U(1)_L$ potential. Thanks to the Yukawa couplings they interact with the left doublet of scalars. They interact also weakly via the $SU(2)$ gauge fields and gravitationally via the metric. However with the mediation of the latter they interact also with the right fermions (right-handed particles and left-handed antiparticles) $\left(\begin{matrix} u_R\\ d_R\end{matrix}\right)$ .

Something more precise can be said about the mixing of the $U(1)_{L/R} $ fields and the third component of the $SU(2)$ gauge field. The $U(1)_L$ group is the group whose generator is  the hypercharge $Y_L= Q_L-T_3$, where $Q$ is the generator whose eigenvalues are the electromagnetic charges and 
$T_3$ is the third generator of $SU(2)$. The two corresponding fields $W_{3\mu}$ and $B^{(-)}_\mu$ combine to form the $A^{(-)}_\mu$ and $Z^{(-)}_\mu$ vector fields which, according to
the standard conventions, are defined by
\be
\left( \begin{matrix} A^{(-)}\\ Z^{(-)}\end{matrix}\right)= \left(\begin{matrix} -\frac {{\mathsf e}_L}{ {\sf g}_B^-} & -\frac {{\mathsf e}_L}{{\sf g}_W}\\ \frac {{\mathsf e}_L}{{\sf g}_W}&  -\frac {{\mathsf e}_L}{ {\sf g}_B^-}\end{matrix}\right)\left(\begin{matrix} B^{(-)}\\W_3\end{matrix}\right) , \quad\quad {\rm with} \quad\quad \left(\frac {{\mathsf e}_L}{ {\sf g}_B^-} \right)^2 + \left( \frac {{\mathsf e}_L}{{\sf g}_W}\right)^2=1,
\label{AandZL}
\ee
where ${\mathsf e}_L$ is of course the ordinary SM electric charge and $A^{(-)}$ the ordinary electromagnetic potential.

In analogy we can introduce $A{(+)}$ and $Z^{(+)}$ fields in the right sector via
\be
\left( \begin{matrix} A^{(+)}\\ Z^{(+)}\end{matrix}\right)= \left(\begin{matrix} -\frac {{\mathsf e}_R}{ {\sf g}_B^+} & -\frac {{\mathsf e}_R}{{\sf g}_W}\\ \frac {{\mathsf e}_R}{{\sf g}_W}&  -\frac {{\mathsf e}_R}{ {\sf g}_B^+}\end{matrix}\right)\left(\begin{matrix} B^{(+)}\\W_3\end{matrix}\right) , \quad\quad {\rm with} \quad\quad \left(\frac {{\mathsf e}_R}{ {\sf g}_B^+} \right)^2 + \left( \frac {{\mathsf e}_R}{{\sf g}_W}\right)^2=1,
\label{AandZR}
\ee
where ${\mathsf e}_R$ and $A^{(+)}$, respectively, are the electric charge and electromagnetic potential in the mirror sector.

These and other things can be said about $\cal T$, but the obvious question is: if the left sector represents the SM in interaction with gravity, what is the interpretation of the right sector? To be less generic, if we would like the left sector to provide a  basic description of the evolution of the very early universe there is certainly something missing in $\cal T$, at least one or more fields that can describe the inflationary period and possibly a quintessence field, if the cosmological constants are not enough to describe the present accelerating expansion. Their action can be easily written down, and we abstain from it only for simplicity. But all we need to meet such cosmological ambitions is there. In particular the left-handed part with the addition of an inflaton field, for instance, can effectively describe, resting on the background of a LFRW metric, the physics of the universe's evolution (inflation, reheating, particle creation and density perturbations, perhaps even dark energy, for reviews see \cite{parker69,parker71,ford77,traschen90,Mukhanov92,shtanov95,kofman95,Greene97,Brandenberger03,
Tsuchikawa03,Vasquez21,Ford21,bambi2021,Kallosh25,Burgess} and references therein). Then the question is: if this is the correct picture, what does the right part represent in it? It does not take much imagination to see in it a candidate for the dark matter. This part (the right one) of the total matter+energy evolves in a way parallel to the ordinary matter+energy, although with different coupling constants, masses and cosmological constant. And at the present energy scale the interaction between the two reduces to the gravitational one, since the common weak force between left and right matter has a range which is too short to be effective at macroscopic distances (unless very high energy scattering phenomena are considered). What makes the difference between the two half theories are the couplings, scalar masses and cosmological constant, beside the handedness of the respective fermions which are opposite in the two sectors. It is likely that the evolution in the right part (the dark matter?) might be sensibly different from the ordinary matter in what concerns inflation, particle creation, density perturbations and so on. Summarizing, the idea is that there is another world (the right one) out there which we cannot `see', but at most `feel' via gravity, and also via the weak force if the energy is high enough. This is suggestive, but it remains for us to explain many aspects, of which the most important is  why the amount of dark matter is more than five times larger than the visible matter. And figure out experiments that permit to falsify the idea. 

The literature on dark matter is vast, with a variety of different ideas and proposals. It is classified in cold, warm and hot dark matter. Focusing on the most popular cold dark matter (CDM), it may be made of baryonic objects, i.e. made of the SM baryons, like MACHOs (massive astrophysical compact halo objects) or primordial black holes. The non-baryonic matter might be made of particles: massive neutrinos, axions, or weakly interacting particles present in supersymmetric models: neutralinos (a mixture of supersymmetric partners), fotinos, Binos,... popularly denoted by the acronim WIMPs  (weakly interacting massive particles). WIMPs have been considered as among the most likely candidates for dark matter. They are massive, thus they feel gravity. They belong to the same left sector as the SM particles, and usually are taken from supersymmetric extensions of the SM, or supergravity, or superstring inspired models. But the essential aspect is that they weakly interact with the SM sector. The fact that they are taken from the SM sector renders WIMPs' construction frankly speaking somewhat contrived. In this sense the right sector, ${\cal T}_R$, sounds a more natural or, at least, less contrived candidate of dark matter. But certainly we have a lot to learn from the WIMP literature, see \cite{bergstrom00,bertone05,arina12,drees12,boucenna,racker14,bambi2021,arbey21} for reviews. In this sense the most striking aspect of it is the so-called `WIMP miracle': with the freeze-out mechanism, WIMPs achieve the relic density for dark matter appropriate to reproduce the latest experimental data, $\Omega_{dm}\approx 0.25$. Although it is not clear what form the mirror matter will take in this new description, whether macroscopic bodies, gas of neutral particles, like the neutrinos in the left sector, or neutral atoms, or all of them together, this `miraculous coincidence' raises some hope. 

Finally if our idea of dark matter had a realistic basis it would unveil a mystery: why does nature use only particles of a precise handedness (with a slogan style abbreviation: left-handed particles and right-handed antiparticles) to build up our universe while disdaining the same particles and antiparticles of opposite handedness? Is it out of style to admit a fascination for this mystery?

\section{Weyl symmetry}

It is often said that whatever local theory we consider, when the energy regime grows very large, masses and dimensionful constants become irrelevant. A related point of view is that these constants may not be true constants, but vacuum expectation values of suitable fields that condense at low enough energies; so that in a fundamental theory only dimensionless constants and fields will appear.  Therefore, since in any such theory there is no explicit scale,  from a conceptual point of view we expect it to be invariant under a rescaling of the metric. From an observational perspective, though, there seems at work a difference of scales when we compare for example the smallness of neutrino masses with those of ordinary particles. Not to speak of the difference in scale between the vacuum energy of the standard model and the observed value of the cosmological constant. 

It is not easy to put together the above two views. The theory presented in the first part of this paper, and represented by the action \eqref{totalaction}, may be of help. One can see that the fermionic action (including the gauge and metric couplings, as well as the Yukawa couplings), the gauge field action and the quartic scalar couplings are invariant under local metric rescalings. Thus they are the same at any scale. What breaks conformal invariance is the Einstein-Hilbert action and the kinetic and mass terms of the scalars. Like in several other examples in physics, the latter terms look like symmetry breaking terms of a more fundamental theory which is fully conformal invariant. They can be regarded as the remnants of terms symmetric like  the other terms in the theory, that is invariant not only under rigid rescalings of the metric but also under local ones, but which, due to some mechanism, have lost the larger symmetry. The sequel is an attempt to illustrate such possibility.

As just explained, the full $\cal T$ theory is not invariant under local Weyl transformations 
\be
g_{\mu\nu} \rightarrow  e^{2\omega} g_{\mu\nu}\label{Weyltransf}
\ee
where $\omega $ is a local parameter, but several pieces thereof are. They are $S_f, S_g$ and $S_Y$ plus the quartic scalar potential terms.. The remaining pieces are not in general Weyl invariant as they contain dimensionful constants. But it is actually very simple to transform a local theory into a Weyl invariant one by adding a new field, $\varphi$, the dilaton. Under the same Weyl rescaling it transforms as $\varphi \rightarrow \varphi+\omega$. The procedure is as follows. Let us start from the Christoffel symbols.
They transform as
\be
\Gamma_{\mu\nu}^\lambda \to \Gamma_{\mu\nu}^\lambda + \delta_\mu^\lambda\, \partial_\nu \omega +  \delta_\nu^\lambda\, \partial_\mu \omega- g_{\mu\nu} g^{\lambda\rho} \partial_\rho\omega\label{Christtransf}
\ee
We can construct Weyl-invariant Christoffel symbols as follows
\be
\widetilde \Gamma_{\mu\nu}^{\lambda} = \Gamma_{\mu\nu}^\lambda - \left(\delta_\mu^\lambda\, \partial_\nu \varphi +  \delta_\nu^\lambda\, \partial_\mu \varphi - g_{\mu\nu} g^{\lambda\rho} \partial_\rho\varphi\right)\label{Christ1}
\ee
We can use these symbols to build the Riemann and Ricci tensor. The latter is
\be
\widetilde R_{\mu\nu} = R_{\mu\nu} +2D_\mu\partial_\nu \varphi+g_{\mu\nu}\square\varphi +2 \partial_\mu \varphi \partial_\nu \varphi-2 g_{\mu\nu} \partial \varphi \cdot \partial\varphi\label{tildeRiccimunu}
\ee
and Ricci scalar is
\be
\widetilde R = R + 6\left(\square\varphi -\partial \varphi\cdot \partial\varphi\right), \label{Rscalar}
\ee
$\widetilde R_{\mu\nu}$ is Weyl invariant, while $\widetilde R \rightarrow e^{-2\omega} \widetilde R$. For the sequel let us remark that if we write $\widetilde g_{\mu\nu} = e^{-2\varphi} g_{\mu\nu}$ we can write
\be
\widetilde R_{\mu\nu} (g) = R_{\mu\nu} (e^{-2\varphi}g)\label{basic}
\ee
where the entry $g$ in the round brackets is a shorthand for the metri $g_{\mu\nu}$.

 Now the recipe is as follows. In the action we replace $R$ with $ \widetilde R$. Then we multiply every dimensionful constant of mass dimension $s$ by the factor $e^{-s\varphi}$. When applied to scalar fields we replace the simple derivatives  $\partial_\mu$ by:
\be
{\sfD}_\mu = \partial_\mu +\partial_\mu \varphi
\ee
The pieces  $S_f, S_g$ and $S_Y$, and the  quartic potential scalar, need not be modified because they are already Weyl invariant. In $\cal T$ we actually need two distinct dilatons $\varphi_\pm$, one for each sector. They behave exactly as the just introduced $\varphi$.

Specifically, for $\cal T$ we have the following modifications. The EH part becomes
\be
S^{(c\pm)}_{EH} =-\frac 1{2\kappa} \int d^4x \sqrt{g}\, e^{-2\varphi_\pm} \left(\widetilde R_\pm + {\mathfrak c}_\pm\, e^{-2\,\varphi_\pm}\right)\label{ScEH}
\ee
where $ \widetilde R_\pm = R + 6\left(\square\varphi_\pm -\partial \varphi_\pm\cdot \partial\varphi_\pm\right)$, 
and the doublet scalar action becomes
\be
S_{d}^{(c\pm)}= \int d^4 x \,\sqrt{g} \left[ g^{\mu\nu} \ED^{\pm}_\mu H_\pm^\dagger \ED^{\pm}_\nu H_\pm -e^{-2\varphi_\pm}  M_\pm^2 H_\pm^\dagger H_\pm -\frac {\lambda_\pm}4 \left(H_\pm^\dagger H_\pm\right)^2\right]\label{Hscalar+-}
\ee
where $\ED^{\pm}_\mu= \partial_\mu + \partial_\mu \varphi_\pm- i\sfg W_\mu=\sfD_\mu- i\sfg W_\mu$ , $W_\mu$ being a gauge field valued in the $SU(2)$ Lie algebra representation  to which $H_\pm$ belongs.

The Weyl invariant generalization of $\cal T$ is therefore
\be
S^{(c)}= S_f+S_g+S_Y + S_{EH}^{(c)} + S_d^{(c)}\label{totalactionc}
\ee
where $ S_{EH}^{(c)}=  S_{EH}^{(c+)}+  S_{EH}^{(c-)}$  and $ S_d^{(c)}= S_d^{(c+)}+ S_d^{(c-)}$.

For later discussion we add also Weyl invariant action terms. One is the higher derivative term 
\be
S _C = \frac 1{\eta}  \int d^4x \sqrt{g}\, C_{\mu\nu\lambda\rho} C^{\mu\nu\lambda\rho}\label{Weyltensoraction}
\ee
$C_{\mu\nu\lambda\rho}$ is the Weyl tensor ($C^{\mu}{}_{\nu\lambda\rho}$ is invariant under Weyl transformations). If we  disregard total derivatives in the action, \eqref{Weyltensoraction} can be replaced by
\be
S'_C =  -\frac 2\eta \int d^4x \sqrt{g} \left( - R_{\mu\nu} R^{\mu\nu}+ \frac 13 R^2\right)  \label{Sc'W}
\ee 
The quadratic terms in brackets contain higher derivative kinetic and interaction terms.
Another similar term can be constructed using the following field
\be
Q_\pm = \square \varphi_\pm -\partial_\mu \varphi_\pm \partial^\mu \varphi_\pm+\frac 16 R\label{exprQ}
\ee
Under a Weyl transformation it transforms as $Q_\pm \rightarrow e^{-2\omega} Q_\pm$. Therefore
\be
S_{Q_\pm}=\frac 6{\gamma_\pm} \int d^4 x\, \sqrt{g}\, Q_\pm^2 \label{intQ2}
\ee
is Weyl-invariant.

The theory defined by \eqref{totalactionc}, with the possible addition of \eqref{Weyltensoraction} and $S_{Q_\pm}$, 
 has the same symmetries as $S$, \eqref{totalaction}. In particular it is invariant under the diffeomorphisms spanned by the parameter $\xi^\mu$, with the dilaton transforming as
\be
\delta \varphi_\pm= \xi ^\mu \partial_\mu \varphi_\pm\label{varphidiff}
\ee
In addition it is conformally invariant. It should be appreciated that conformal invariance of the action $S^{(c)}$ changes the nature of constants and masses. The latter are dressed by a dilaton factor, they are actually transformed into fields. For instance the mass factor $ M_\pm e^{-2\varphi_\pm}$, and other similar factors, can take any value, from 0 to $\infty$, without changing the value of the action. Let us call the new theory $\cal TW$. Fron  now on, however,  we will drop the $\pm$ labels as the two halves of $\cal T$ can be dealt with independently (as far as the topics covered in the sequel are concerned).
\vskip 0.2cm
Refs.:\cite{ghilencea21,ghilencea23,ghilencea24,mohammedi2024,roumelioti24,ghilencea26}, see also the reviews \cite{scholz2018,rachwal2022}.

\subsection{Why Weyl invariance?}

As pointed out above, enlarging the symmetry to the maximal one allowed in a model, is a time-honored practice to construct a more fundamental theory. There may be also other motivations which have to do with the specific symmetry in question, i.e. Weyl symmetry. They have been presented in a previous paper, \cite{bonora25}, of which a short summary is given hereafter. 

The first question considered there was the search for a classical solution in a simplified conformal invariant model of the form
\be
S_1^{(c)}=-\frac 1{2\kappa} \int d^4x \sqrt{g}\, e^{-2\varphi} \left(\widetilde R + {\mathfrak c}_\pm\, e^{-2\,\varphi}\right)+ \frac 12  \int d^4 x \,\sqrt{g} \left[ g^{\mu\nu} \sfD_\mu \Phi \sfD_\nu \Phi -e^{-2\varphi}  m^2 \Phi^2 -\frac {\lambda}4\Phi^4\right]\label{Strunc}
\ee
assuming the metric is the Friedmann-Robertson one. The solution exists and has the form
\be
\Phi(t)=\frac {\alpha}t, \quad\quad \varphi(t) = \ln \left(\beta t\right), \quad\quad a(t)= \gamma\, t\label{ansatz}
\ee
with cosmological parameters 
\be
\rho \sim P \sim \frac 1{a^4}\label{rhoP}
\ee
This solution has been proposed in \cite{bonora25} as fit to describe a universe evolution prior to the period of inflation. On the backdrop of this solution the following interpretation of the cosmological constant problem was suggested.

The vacuum energy density due to gravitation is represented by
\be
\rho_\Lambda = \frac {\mathfrak c}{2\kappa} \label{cosmdensity}
\ee
The observed value is
\be
|\rho_\Lambda^{(obs)}|\sim 2 \times 10^{-10}erg/cm^3\label{Obslambda}
\ee
The problem is that putting together in a unique theory gravity and matter, the matter field theory comes with its own vacuum energy. The latter is always a divergent quantity and can be estimated only using different cutoffs, \cite{Gellmann68,shifman79,carroll}. If the QCD scale is used the result is $\rho_{vac}^{QCD} \sim 1.6\times10^{36} erg/cm^3$. With the electroweak scale it is $\rho_{vac}^{EW} \sim 3\times 10^{47} erg/cm^3$ and if the cutoff is the Planck mass, one gets $\rho_{vac}^{Pl} \sim 2\times 10^{110} erg/cm^3$. In any case the gap with \eqref{Obslambda} is gigantic and utterly unnatural, see \cite{weinberg89,weinberg00,vilenkin01} and references therein. Proceeding in the quantization without solving this paradox, or at least without hinting at a possible solution, is simply ridiculous. But if we look instead at $S^{(c)}$, \eqref{totalactionc}, the problem can appear in a different light. First,  the fermion and gauge parts of the action, as well as the Yukawa coupling and the quartic scalar couplings, are unaffected by Weyl transformations. On the contrary the other terms in $S^{(c)}$, and in particular the cosmological constant term undergo drastic changes under a Weyl transformation.  The `gauge' $\varphi=0$ reproduces the just mentioned unnatural situation, but if we choose a sufficiently negative value for $\varphi$, for instance a `gauge' $\varphi\approx -25,\quad e^{-4\varphi}\sim  10^{43}$, or a similar one, the effective cosmological constant takes on a value for which the gravitational vacuum energy may become comparable with the value of the vacuum energy of the theory, whatever it may be.  

The difficulty in nowadays quantum field theory is that we are able to quantize field theories only via a perturbative series. Therefore, for instance, the smallness of the measured cosmological constant disappears compared to the quantum corrections of the SM. Simply it does not make much sense to juxtapose matter and gravity (if the cosmological constant represents its vacuum energy) in the same quantum theory. However, the theory $\cal TW$ is conformal invariant. Therefore we can quantize it at the scale (i.e. the `gauge') where the perturbative approach makes sense, and transfer the quantized results (renormalization and unitarity) to the other scales, much as we do in quantum gauge theories where we quantize in our favorite gauge and then we prove gauge fixing independence. In the present case therefore what we have to do next is to do quantization and show that conformal invariance is preserved by quantization. However before passing to quantization a short mention is in order concerning the mechanism that determine a constant `gauge' the dilaton. In \cite{bonora25} it was suggested that it should be similar to the spin condensation in ferromagnetism, whereby to a disordered system of spins at high temperature there succeeds a system of patches of spins oriented in the same direction when temperature decreases.  In the present case (probably several) dilaton fields get stuck at constant values within the same patches when the universe temperature starts dropping after the big-bang.

\section{Quantization}

As shown above it is relatively easy to transform a classical local field theory containing matter and gravity into a conformally invariant one. The price is cheap, it is enough to introduce a dilaton field and suitably manipulate with it the terms which are not  by themselves confomal invariant, \cite{codello12}. The main question is whether the invariance survives the quantization process. Quantization, at least perturbative quantization, requires that propagators and vertices be unambiguously defined. This is the case for $\cal TW$ because the theory has been constructed in order to guarantee this requirement, namely absence of O-type anomalies. The next requirement for a quantum theory to be consistent is unitarity and renormalizability. 

Concerning renormalizability, if the gravitational sector of the theory is limited to the EH action we meet immediately an insurmountable difficulty. For the graviton propagator extracted
from the EH action after gauge fixing is inversely quadratic in the momentum in the UV limit, while, on the other hand, it contains an infinite number of interaction terms. This implies that the UV divergence of the Feynman diagrams increases indefinitely with the number of loops, rendering the theory hopelessly non-renormalizable. This flaw can be repaired by introducing quadratic terms in the curvature, \cite{utiyama62,deser75,gavrielides75,stelle1977}. The new graviton propagator, after suitably gauge fixing, is now UV asymptotically quartic in momentum. This prevents the uncontrollable increase of UV divergences and transforms the theory into a power-counting renormalizable one. It requires an infinite number of renormalition constant, but they are all, except a small number, related to the graviton field redefinition, therefore non-physical.  

The previous modification, however, triggers the appearance of a new entry. We can get an idea of how this works as follows. If we limit ourselves to the lowest order kinetic operator for the graviton $h_{\mu\nu}= g_{\mu\nu}-\eta_{\mu\nu}$ in $S_{EH}$ we find, after a suitable gauge fixing, $\alpha \square$, where $\alpha$ is a constant with the dimension of a square mass. The addition of the term \eqref{Weyltensoraction} or \eqref{intQ2} brings in the kinetic operator quartic derivatives, which in the simplest case can be represented as $\beta \square^2$, where $\beta$ is a dimensionless constant. Disregarding the tensorial factors, the propagator is proportional to  the inverse of $\alpha \square + \beta \square^2$ , i.e. the inverse of $-\alpha p^2 + \beta p^4$, which can be written as follows
\be
\frac 1{-\alpha p^2 + \beta p^4}= \frac 1{p^2 (-\alpha +\beta p^2)} = -\frac 1{\alpha} \left( \frac 1{p^2}- \frac 1{p^2-\frac {\alpha}{\beta}}\right)\label{propghmunu}
\ee
This inevitably introduces a quadratic pole with negative residue, corresponding to a negative norm state, which is likely to violate unitarity.

The occurrence of physical ghosts in similar gravity or gravity plus matter theories has been confirmed in several papers, \cite{stelle1977,julve-tonin1978,anselmi17,salvio18,strumia19,sibold21,donogue22,oda22a,oda22b,oda22c,kubo23,sibold23,oda23,holdom24,sibold24,oda24,oda24b,buoninfante25}. Based on these results one can reasonably expect that the theory $\cal TW$, defined by the action $S^{(c)}$,  eq.\eqref{totalactionc}, may present problems both for unitarity and renormalizability. Renormalization requires the addition of quadratic curvature terms to the action, on the other hand this may put in jeopardy unitarity. This conflict is still unsolved although some progress may be at hand, see \cite{tkach18,sibold21,sibold23,oda26,shapiro26}. Inspired by these papers we would like to see whether algebraic renormalization \cite{BRS76,piguetsorella} with the BPHZL subtraction scheme, \cite{zimmer68,zimmer69,zimmer73,bogo59,Lowen75a,Lowen75b,Lowen76}, is applicable to a theory like $\cal TW$. To keep the notation and formulas within manageable limits we shall deal not with $\cal TW$, but with a simpler action (in particular we drop the $_\pm$ labels and the matter and gauge fields), which, however, we believe contains the essential aspects of the problem. The starting action will be
\be
S_0 =S_{EH}^{(c)} + S_C+S_Q= -\frac 1{2\kappa} \int d^4x \sqrt{g}\, e^{-2\varphi} \widetilde R-\frac 2\eta \int d^4x \sqrt{g} \left( - R_{\mu\nu} R^{\mu\nu}+ \frac 13 R^2\right) + \frac 6\gamma \int d^4x  \sqrt{g}\, Q^2 \label{S0}
\ee
with ${\mathfrak c}=0$. $S_0$ is conformal invariant. 
Had we inserted the very EH action instead the of first term on the right, we would have broken conformal invariance. But we wish to preserve it and quantize it in this invariant form. Quantizing will require breaking both diffeomorphism and conformal symmetry with specific gauge fixing term(s) with the aim of recovering them in the form of BRST symmetries. 
Let us remark that the action \eqref{S0} is the most general diffeomorphism and conformal invariant action containing two and four derivatives of the graviton and the dilaton. This point will be made clear later on when studying the BRST cohomology 
of the theory.

The first task to be complied with in order to quantize $S_0$ is to determine the graviton and dilaton propagtors.

\subsection{The propagator}

The quadratic terms of the various pieces above are
\be
S_{EH}^{(c, quad)}&\!=\!&-\frac 1{2\kappa}\int d^4x \, \left[ \frac 14 h^{\mu\nu} \square h_{\mu\nu}- \frac 14 h\square h +\frac 12 \partial_\lambda h^{\mu\lambda}\partial^\nu h_{\mu\lambda}- \frac 12 \partial_\lambda  h^{\mu\lambda} \partial_\lambda h\right]\0\\
&&+\frac 1{2\kappa}\int d^4x \,\left[ 6 \varphi \square \varphi+2 h^{\mu\nu} \left( \partial_\mu\partial_\nu-\eta_{\mu\nu} \square \right)\varphi\right],\label{quadSEHc}
\ee 
where $h=h^\mu_\mu$. The quadratic $h_{\mu\nu}$ kinetic operator in momentum  space is
\be
\widetilde K^{(EH)}_{\mu\nu,\lambda\rho} = \frac {p^2}{8\kappa} \left(P^{(2)} -2P^{(0)} \right)_{\mu\nu,\lambda\rho}. \label{KEH}
\ee
 The projectors $P^{(2)}$ and  $P^{(0)}$ as well as others are given in Appendix A.

Moreover
\be
S_C^{quad} = -\frac 2\eta \int d^4x \, h^{\mu\nu}\left[-\frac 14 \square ^2 \eta_{\mu\lambda} \eta_{\nu\rho} +\frac 1{12}\square^2 \eta_{\mu\nu} \eta_{\lambda\rho} -\frac 16 \partial_\mu\partial_\nu\partial_\lambda\partial_\rho +\frac 12\eta_{\mu\lambda} \square  \partial_\nu\partial_\rho-\frac 16\eta_{\mu\nu} \square  \partial_\lambda\partial_\rho \right]h^{\lambda\rho},
\label{SCquad}
\ee
with graviton quadratic kinetic operator
\be
\widetilde K^{(C)}_{\mu\nu,\lambda\rho} = \frac {p^4}{2\eta} P^{(2)}_{\mu\nu,\lambda\rho}\label{KC}
\ee
and, finally,
\be
S_Q^{(quad)} &=&\frac 1\gamma \int d^4x \, h^{\mu\nu}\left[\frac 16 \partial_\mu\partial_\nu\partial_\lambda\partial_\rho -\frac 13 \eta_{\mu\lambda} \square  \partial_\nu\partial_\rho+ \frac 16 \square^2 \eta_{\mu\nu} \eta_{\lambda\rho}\right] h^{\lambda\rho}\0\\
&& +\frac 2\gamma \int d^4x \, \left[ 3 \varphi \square^2 \varphi +\varphi \square\left(\partial_\mu\partial_\nu -\eta_{\mu\nu}\square\right) h^{\mu\nu} \right]\label{SQquad}
\ee 
The corresponding graviton-graviton kinetic operator is 
\be
\widetilde K^{(Q)}_{\mu\nu,\lambda\rho}=  \frac {p^4}{2\gamma} P^{(0)}_{\mu\nu,\lambda\rho}\label{KQ}
\ee
It is evident that we cannot invert the kinetic operator for $h_{\mu\nu}$ alone, but we have to consider the joint kinetic operator for $h_{\mu\nu}$ and $\varphi$. It is not surprising that the latter is not invertible because of zero modes due to gauge invariance. We have to fix the gauge.
\subsubsection{Gauge fixing}

In an ordinary gravitational theory the Slavnov-Taylor (ST)  identity is the functional expression that represents its invariance under diffeomorphisms. To quantize the theory one has to fix the gauge and recover it as a BRST symmetry of the quantum action. The ST identity is precisely a manifestation of such a symmetry. In the present context the theory has an additional symmetry, under conformal transformations. It would seem natural and more manageable to break diffeomorphism symmetry without breaking conformal invariance and subsequently break the residual conformal invariance without breaking anew diffeomorphism invariance. All attempts in this direction have failed: breaking diffeomorphism invariance without breaking conformal invariance is possible but would lead to propagators unfit to satisfy the Lowenstein conditions (see below). On the other hand one single breaking term turns out not to be enough to guarantee the existence of a propagator, even though it accidentally breaks conformal invariance (a higher order breaking). In order to make sense of quantization (existence of propagators) one has to add an additional (hard) conformal symmetry breaking term.

The gist of this discussion is that one cannot disentangle conformal and diffeomorphism   symmetry and quantize them separately. 
After all this is not so surprising because the diffeomorphisms and conformal transformations form not a direct product group but a semidirect product one. Let us call it $\cal G$; it includes an action of diffeomorphisms over the conformal transformation parameter represented by $\delta_\xi \omega = \xi\! \cdot \!\partial \omega$. It is clear that we have to construct the relevant ST identity for such overall symmetry $\cal G$. We introduce the familiar BRST transformation symbol $\mathsf s$ promoting $\xi^\mu $ and $\omega$ to anticommuting fields and setting
\be
\mathsf s= \delta_\xi + \delta_\omega, \quad\quad {\mathsf s}^2=0\label{ess}
\ee
with 
\be
\mathsf s\, g_{\mu\nu} &=&(\delta_\xi + \delta_\omega) g_{\mu\nu} = \xi^\lambda \partial_\lambda g_{\mu\nu}+ \partial_\mu \xi^\lambda g_{\lambda\nu} + \partial_\nu \xi^\lambda g_{\lambda\nu} + 2 \omega g_{\mu\nu}\0\\
\mathsf s\, \varphi&=&(\delta_\xi + \delta_\omega)\varphi= \xi^\lambda\partial_\lambda \varphi + \omega\0\\
 \mathsf s \,\xi^\mu&=&(\delta_\xi + \delta_\omega)\xi^\mu= \xi^\lambda \partial_\lambda \xi^\mu\0\\
  \mathsf s\, \omega&=&(\delta_\xi + \delta_\omega)\omega=\xi^\lambda \partial_\lambda \omega\label{sBRST}
  \ee 
It is easy to prove that ${\mathsf s}^2=0$. Remember that $g_{\mu\nu}= \eta_{\mu\nu}+h_{\mu\nu}$.

One interesting possibility is the gauge-fixing term
\be
S_{g.f.}^{(diff)}= -\frac 1{2\sqrt \kappa} \int d^4x \,g_{\mu\nu}\,\left( \partial^\mu b^\nu + \partial ^\nu b^\mu \right) - \frac {\alpha_0}2  \int d^4x\,\eta_{\mu\nu} b^\mu b^\nu \label{gfdiff}
\ee
where $\partial^\mu = \eta^{\mu\nu} \partial_\nu$.
This is essentially the gauge fixing term used by \cite{sibold21}. It breaks not only the diffeomorphism invariance but also conformal invariance. However it is not enough to allow us to invert the kinetic operator. We proceed as follows. Instead of \eqref{gfdiff} we use
\be
S_{g.f.}^{(diff,c)}= -\frac 1{2\sqrt \kappa} \int d^4x \left(g_{\mu\nu}- 2 \eta_{\mu\nu} \varphi\right)\, \left( \partial^\mu b^\nu + \partial ^\nu b^\mu\right) - \frac {\alpha_0}2 \int d^4x\, \eta_{\mu\nu} b^\mu b^\nu \label{gfdiffconf}
\ee
which breaks diffeomorphisms while preserving conformal invariance at lowest order. This gauge fixing gives rise to the FP term
\be
S_{FP}=- \frac 1{2\sqrt \kappa} \int d^4x\,\left(\left(\partial^\mu {\overline \xi}^\nu + \partial ^\nu \overline \xi^\mu\right) \mathsf s \left(g_{\mu\nu}-2\eta_{\mu\nu}\varphi\right) \right) \label{FPdiff}
\ee 
where the antighost field $\xi^\mu$ transform as
\be
\mathsf s\, {\bar \xi}^\mu = -b^\mu,\quad\quad
\mathsf s\, b_\mu&=&0\label{sBRST2}
\ee

The quadratic derivative version of \eqref{S0} and \eqref{gfdiffconf} shows that one cannot disentangle the kinetic quadratic terms for $h_{\mu\nu}, b^\mu$ and $\varphi$, but we have to consider the triple joint kinetic term $h_{\mu\nu} - b^\mu - \varphi $. But even with the gauge fixing \eqref{gfdiffconf} this joint kinetic term is not invertible (the kinetic operator contains zero modes). As a consequence we have to fix the residual conformal gauge symmetry. The relevant gauge fixing term we choose is:
\be
S_{g.f.}^{(c)} = -\int d^4x\, \left( b\, {\square}\varphi- \frac {\beta_0}2 b^2\right) \label{gfconf}
\ee
The corresponding FP term is 
\be
S_{FP}^{(c)} = -\int d^4x\,  \overline\omega\,\mathsf s\, (\square \varphi) \label{FPconf}
\ee
In both formulas $\square$ is the ordinary D'Alambertian.

To complete the set of BRST transformations \eqref{sBRST} and \eqref{sBRST2} we assume the following transformations
\be
 \mathsf s\,\overline \omega =-b, \quad\quad \mathsf s\, b=0. \label{sbaromega}
\ee
Of course, everywhere,  $\mathsf s ^2=0$. The quadratic terms \eqref{gfconf} and \eqref{FPconf}, from which the relevant propagators have to be extracted, are
\be
S_{g.f.}^{(c, quad)} = \int d^4x\, \left(\partial_\mu b\, \partial^\mu \varphi+ \frac {\beta_0}2 b^2\right), \quad\quad {\rm and} \quad\quad S_{FP}^{(c,quad)} =- \int d^4x \, \overline\omega\, \partial_\mu\partial ^\mu \omega\label{quadgfFP}
\ee

It is worth stressing that the diffeomorphism symmetry breaking term $S_{g.f}^{(diff,c)}$ and the corresponding FP term are conformal invariant to lowest order. This will avoid the appearance of a mixing $\overline \xi^\mu - \omega$ in the ghost  kinetic term, which induces a considerable simplification in the sequel.  

Before proceeding further we collect the various action pieces under a unique symbol
\be 
S= S_0 + S_{g.f.}^{(diff,c)} + S_{g.f.}^{(c)} + S_{FP}^{(diff)}+S_{FP}^{(c)}\label{S}
\ee
from eqs.(\ref{S0},\ref{gfdiffconf},\ref{FPdiff},\ref{gfconf},\ref{FPconf}).

\subsubsection{Kinetic operator}

We can now write down the full kinetic operator, which consists of the three graviton-graviton contributions above. But we have also the $h_{\mu\nu} - b_\alpha$, the $h_{\mu\nu} - \varphi$ and the $b_\alpha- \varphi$ terms, given,respectively, by
\be
\widetilde L_{\mu\nu,\alpha} =-\frac i{4\sqrt \kappa} \left(p_\mu \eta_{\nu\alpha} + p_\nu \eta_{\mu\alpha}\right), \quad\quad \widetilde\ell_{\mu\nu}= \left(\frac {p^4}\gamma- \frac {p^2}{2\kappa} \right) \theta_{\mu\nu},\quad\quad \widetilde t_\alpha= \frac {2i}{\sqrt \kappa} p_\alpha.\label{tildeLtildeell}
\ee 
Moreover we have the $b_\alpha-b_\beta$, the $b-b$, the $\varphi-\varphi $ and the $b-\varphi$ pieces given, respectively, by
\be
\widetilde Q_{\alpha\beta}= -\frac {\alpha_0}2 \eta_{\alpha\beta}, \quad\quad \widetilde Q_{\beta_0}=\frac {\beta_0}2, \quad\quad \widetilde Q= 6 \left(\frac {p^4}\gamma- \frac {p^2}{2\kappa} \right), \quad\quad \widetilde V = p^2\label{QabQV}
\ee
Altogether the kinetic operator in momentum space can be represented by the matrix
\be
\widetilde{\cal K}=  \left( 
\begin{matrix}\widetilde K^{\mu\nu,\lambda\rho} & \widetilde L^{\mu\nu,\beta} &\widetilde \ell^{\mu\nu} &0\\
-\widetilde  L^{\lambda\rho, \alpha}& \widetilde Q^{\alpha\beta}&\widetilde t^\alpha&0\\
\widetilde \ell^{\lambda\rho} & -\widetilde t^\beta & \widetilde Q & \widetilde  V\\
0&0& \widetilde  V & \widetilde Q_{\beta_0}\end{matrix} \right)\quad\quad{\rm to\, be\,applied\, to\, the\, vector} \quad\quad  \left( \begin{matrix}\widetilde h_{\lambda\rho}\\\widetilde b_\beta \\ \widetilde\varphi\\\widetilde b\end{matrix} \right)
\ee
of Fourier-transformed fields. From eqs.(\ref{KEH},\ref{KQ},\ref{KC}), see Appendix A for the $P^{(2)}$ and $P^{(0)}$ symbols,
\be 
\widetilde K_{\mu\nu,\lambda\rho} = \left(\left(\frac {p^2}{8\kappa}+ \frac {p^4}{2\eta} \right)P^{(2)} +\left(\frac {p^4}{2\gamma}- \frac {p^2}{4\kappa} \right)P^{(0)}\right)_{\mu\nu\lambda\rho}\label{Kmnlr}
\ee
The inverse of $\widetilde{\cal K}$ has the structure
\be
\widetilde{\cal P}=\left( \begin{matrix} \widetilde P_{\lambda\rho,\sigma\tau} &\widetilde R_{\lambda\rho,\gamma} & \widetilde v_{\lambda\rho}& \widetilde u_{\lambda\rho}\\
-\widetilde R_{\sigma\tau,\beta}& \widetilde Z_{\beta\gamma}& \widetilde v_\beta&\widetilde u_\beta\\
\widetilde v_{\sigma\tau} &- \widetilde v_\gamma& \widetilde Y& \widetilde X\\
\widetilde u_{\sigma\tau} & - \widetilde u_\gamma & \widetilde X&\widetilde  T \end{matrix} \right)\label{calP}
\ee
where
\be
\widetilde P_{\mu\nu,\lambda\rho }&\!\!=\!\!&\left(A P^{(2)} +B P^{(1)}+C P^{(0)}+D \overline P^{(0)} + E T^{(0)}\right)_{\mu\nu,\lambda\rho},\0\\
\widetilde R _{\mu\nu,\alpha} &=& i\left( F \,\omega_{\mu\nu}p_\alpha +G\, (p_\mu \eta_{\nu\alpha} +p_\nu \eta_{\mu\alpha}) +K\, \eta_{\mu\nu} p_\alpha\right),\0\\
\widetilde v_{\mu\nu} &=& H\,\theta_{\mu\nu} +L\, \omega_{\mu\nu}, \quad\quad \widetilde u_{\mu\nu}= N\,\theta_{\mu\nu} + P\, \omega_{\mu\nu} ,\0\\ 
\widetilde v_\mu& =& i M p_\mu, \quad\quad \widetilde u_\mu = i R\, p_\mu,\0 \\
\widetilde Z_{\alpha\beta} &=&Z\theta_{\alpha\beta} +W \omega_{\alpha\beta} \label{ABCDE}
\ee
$A,B,C,D,E,F,G,H,K,L,M,N,P,R,Z,W$ as well as $\widetilde X,\widetilde Y, \widetilde T$ are functions of the momentum to be determined (see Appendix A for the symbols in the first line).

\subsubsection{Propagators}

First of all it is easy to check that the matrix $\widetilde{\cal K}$ does not possess zero modes, so that its inversion is guaranteed.
To invert $\widetilde{\cal K}$ we must check that the matrix product  $\widetilde{\cal K} \,\widetilde {\cal P}$ yields the identity represented by the four component vector 
$\left(\begin{matrix}\frac 12\left(\delta^\mu_ \sigma\delta^\nu_\tau + \delta^\nu _\sigma\delta ^\mu_\tau\right)\\ \delta^\alpha_\gamma \\ 1\\ 1\end{matrix}\right)$. This is a lengthy but straightforward operation, which leads to:
\be 
&&A= \frac 1{\frac {p^2}{8\kappa} +  \frac {p^4}{2\eta}}, \quad\quad B= -\frac {4\alpha_0\kappa}{p^2}, \quad \quad C=\frac 2{\frac {p^4}\gamma- \frac {p^2}{2\kappa} }-\frac {6\beta_0}{p^4},\quad  D= \frac {4\beta_0}{p^4}+\frac {2\alpha_0\kappa}{p^2}, \quad E=- \frac {4\sqrt 3 \beta_0}{p^2},\0\\
&&F=-G = \frac {2\sqrt\kappa}{p^2}, \quad\quad H=\frac {\beta_0}{p^4}, \quad L=  \frac {2\beta_0}{p^4},\quad\quad N=-\frac {2}{p^2}, \quad P= - \frac {4}{p^2},\0\\
&&\widetilde X= \frac 1 {p^2},\quad\quad \widetilde Y=\frac {\beta_0}{2p^4},\\
&&K=M=R=\widetilde T=Z=W=0.\0\\\label{ABCDEFGH} 
\ee
From this we can read off the propagators in momentum space. We start with the graviton-graviton $\langle h_{\mu\nu}h_{\lambda\rho}\rangle$  propagators (we drop for simplicity the Lorentz labels)
\be
&&\langle h\,h\rangle^{(2)}=  \frac i{\frac {p^2}{8\kappa} +  \frac {p^4}{2\eta}}P^{(2)}, \quad\quad \langle h\,h\rangle^{(1)}= -i \frac {4\alpha_0\kappa}{p^2} P^{(1)}, \quad\quad \langle h\,h\rangle^{(0)}=i \left(\frac 2{\frac {p^4}\gamma- \frac {p^2}{2\kappa} }-\frac {6\beta_0}{p^4}\right)P^{(0)},\0\\
&& \langle h\,h\rangle^{\overline{( 0)}} =i \left( \frac {4\beta_0}{p^4}+\frac {2\alpha_0\kappa}{p^2}\right)\overline P^{{(0)}} , \quad\quad \langle h\,h\rangle^{(T)}=-i\frac {4\sqrt 3 \beta_0}{p^2}T^{(0)}\label{prophh}
\ee
The remaining propagators are
\be
&&\langle h_{\mu\nu} b_{\alpha} \rangle = \frac {2\sqrt{\kappa}}{p^2} \left(p_\mu \eta_{\nu\alpha} +p_\nu \eta_{\mu\alpha}-\omega_{\mu\nu} p_\alpha\right),\quad\quad \langle h_{\mu\nu}\,\varphi\rangle = i\frac {\beta_0}{p^4} \left(\theta_{\mu\nu} +2\omega_{\mu\nu} \right)\0\\ &&\langle h_{\mu\nu}\,b\rangle= -\frac{2i}{p^2}   \left(\theta_{\mu\nu} +2\omega_{\mu\nu} \right)\label{prophmnb}
\ee
and
\be
&&\langle \varphi\,\varphi\rangle =- i \frac {\beta_0} {2p^4}, \quad\quad \langle b\,\varphi\rangle = \frac i{p^2} , \quad\quad \langle b_\alpha \,b_\beta\rangle =0, \quad\quad \langle b_\alpha \, \varphi \rangle =0 \0\\
&& \langle b_\alpha \,b\rangle =0, \quad\quad \langle b \,b\rangle =0\label{varphivarphi}
\ee

The quadratic sector for the ghosts is separate from the one just discussed, and the $\xi - \bar \xi $ is separate from the $\overline\omega  -\omega$ one. The derivation of the  relevant propagators is very simple
\be
\langle \overline\xi_\mu ,\xi_\nu\rangle = -i \frac {\sqrt \kappa} {p^2} \left(\theta_{\mu\nu} +\frac 12\omega_{\mu\nu}\right), \quad\quad\langle \overline \omega \, \omega\rangle = \frac i{p^2} \label{ghostprop}
\ee

\subsection{BPHZL  and Lowenstein conditions}

Our approach to renormalization is based on the BPHZL subtraction scheme, which is complicated in our case because of the presence of massless states. In other words we need subtractions both in the UV and in the IR. Let a Feynman diagram $\gamma$ be given whose amplitude
\be
J_\gamma(p) = \int  \prod_{l=1}^\ell d^4k_l\, I_\gamma (p,k), \label{Jgamma}
\ee
where $p$ represent a set of external momenta and $k$ a set of internal (loop) momenta, is UV or IR divergent. The integrand is a rational function of the internal and external momenta. The scheme consists in replacing the divergent integral with a well-defined one by subtracting a suitable Taylor expansion in the external momenta to the function $I_\gamma (p,k)$:
\be
\widehat J_\gamma(p)=  \int \prod_{l=1}^\ell d^4k_l\, R_\gamma (p,k)\label{hatJgamma}
\ee
where 
\be
R_\gamma (p,k)= (1-\tau_\gamma) I_\gamma (p,k), \quad\quad\quad (1-\tau_\gamma) =\left(1-t^{\rho(\gamma)-1}_{p^\gamma(s^\gamma-1)}\right) \left(1- t^{\delta(\gamma)} _{p^\gamma s^\gamma} \right)\label{Rgammapk}
\ee
and 
\be
t^{\alpha}_p f(p_1,\ldots, p_n) =\sum_{l=0}^\alpha \frac 1{l!} p^{\mu_1}_{i_1} \ldots p^{\mu_l}_{i_l} \left( \frac {\partial}{\partial p^{\mu_1}_{i_1}}\dots  \frac {\partial}{\partial p^{\mu_l}_{i_l}}f(p_1,\ldots p_n) \right)\Bigg{\vert} _{p_i=0}\label{talpha}
\ee
In this formula $\delta(\gamma)$ and $\rho(\gamma)$ are the UV and IR subtraction degrees of $\gamma$, which we shall choose equal to 4. $s^\gamma$ is a set of positive numbers $s_i$ which accompany divergent external momenta $p_i$ in the following way: any quadratic $p_i^2$ either in vertices or in poles $p_i^{-2}$ in propagators of $\gamma$, are replaced by $p_i^2 -M^2(s-1)^2$. UV subtractions are  implemented at $p_i=s_i=0$, while IR are at $p_i=0,s_i=1$. It is evident that this shift in the square momenta will introduce new terms in the vertices of the action, that have to be taken care of. Another accessory of this approach is the $i\epsilon$ prescription. Any denominator of the form $(q^2- m^2)^n$ is replaces by $(q^2 -m^2 +i\epsilon ({\bf q}^2 +m^2))^n$, where $q$ is a linear combination of internal and external momenta.

If the diagram $\gamma$ contains overlapping lines the formula \eqref{Rgammapk} has to be replaced by the forest formula, \cite{Lowen75a, Lowen75b,Lowen76}, which here will be understood.

In order to be allowed to apply the BPHZL procedure for renormalization we have to show that the requested conditions on the UV and IR behavior of the 1PI amplitudes constructed out of  Feynman diagrams are satisfied. To verify this we collect hereunder the UV and IR divergence degree of  
the propagators in powers of the momentum, denoting the UV power with a top bar and the IR with a bottom one. We consider only the gauge independent propagators, i.e. those that do not depend on $\alpha_0$ and $\beta_0$. We have
\be
&&\overline{\langle h\,h\rangle}^{(2)}=-4, \quad\quad \underline{\langle h\,h\rangle}^{(2)}=-2, \quad\quad
\overline{\langle h\,h\rangle}^{(0)}=-4, \quad\quad \underline{\langle h\,h\rangle}^{(0)}=-2\0\\
&&\overline{\langle b \varphi\rangle} = -2, \quad\quad \underline{\langle b \varphi\rangle} = -2\quad\quad \overline{\langle h_{\mu\nu}\,b\rangle}= -2 , \quad\quad \underline{\langle h_{\mu\nu}\,b\rangle}= -2 \0\\
&& \overline{\langle h_{\mu\nu} b_{\alpha} \rangle}= -1, \quad\quad \underline{\langle h_{\mu\nu} b_{\alpha} \rangle}= -1. \label{UV&IR}
\ee
The last line is however irrelevant for the subsequent discussion because the field $b_\alpha$ has no vertices to attach to in any 1PI graph, since in $S_{g.f.}^{(diff)}$, \eqref{gfdiff}, there are no vertices. For the ghosts we have
\be
\overline{\langle \overline \xi\, \xi\rangle}=-2, \quad\quad 
\underline{\langle \overline \xi\, \xi\rangle}=-2, \quad\quad 
\overline{\langle \overline \omega\, \omega\rangle}=-2, \quad\quad 
\underline{\langle \overline \omega\, \omega\rangle}=-2,\label{UVIRghosts}
\ee

The second ingredient of Feynman graphs are vertices. Their UV and IR behavior, with the same notation as above are
\be
\overline {V_{EH}}=2, \quad \underline {V_{EH}}=2, \quad \overline {V_{C}}=4, \quad \underline {V_{C}}=4,\quad
\overline {V_{Q}}=4, \quad\underline {V_{Q}}=4,\label{gravvert}
\ee
which are the multigraviton vertices present in $S_{EH}^{(c)}, S_{C}$ and $S_Q$, respectively.
Then we have the vertices where the dilaton field enter
\be
\overline{V_{(4)\varphi h}}=4,\quad\quad \underline{V_{(4)\varphi h}}=4,\quad\quad \overline{V_{(2)\varphi h}}=2\quad\quad\underline{V_{(2)\varphi h}}=2,\label{Vvarphi}
\ee 
where $h$ here stands for $h_{\mu\nu}$. The number 4 and 2 in brackets refer to the number of derivatives contained in each vertex. The four derivative vertices come from $S_C$ and $S_Q$, the two derivative ones are in $S_{EH}^{(c)}$. These vertices may contain any number of $h_{\mu\nu}$ and $\varphi$, larger than two. 
 For the ghosts we have
\be
 \overline{V_{\varphi \overline\xi \xi}}=2, \quad\quad\underline{V_{\varphi \overline\xi \xi}}=2, \quad\quad \overline{V_{ \overline\xi \xi h}}=2,  \quad\quad\underline{V_{\overline\xi \xi h}}=2,\quad\quad \overline{V_{h \bar\xi \omega}}=1, \quad\quad  \underline{V_{h \bar\xi \omega}}=1. \label{ghostvert}
 \ee
The label $_{h}$ stands for graviton and the labels $_{\overline \xi, \xi}$ for $\overline \xi_\alpha, \, \xi_\alpha$.

Now let $\gamma$ be  a 1PI Feynman diagram. Its UV divergence is given by
\be
d(\gamma) = 4\ell(\gamma) +\sum_{V\in \gamma} n(\overline V)+ \sum_{I\in \gamma} n(\overline I) \label{UVdivgamma1}
\ee
where $\ell(\gamma), n(V)$ and $n(I)$ are the numbers of loops, of vertices and internal lines (propagators) of $\gamma$. Analogously the IR divergence of $\gamma$ is 
\be
r(\gamma) = 4\ell(\gamma) +\sum_{V\in \gamma} n(\underline V)+ \sum_{I\in \gamma} n(\underline I)\label{IRdivgamma1}
\ee

The number of loops obeys the topological relation
\be
\ell= I-V+1\label{toprel}
\ee
Replacing it in \eqref{UVdivgamma1} and simplifying we get
\be
&&d(\gamma) =4 -2 n(V_{EH}) -2nV_{(2)h\varphi}-2 n(V_{\overline\xi \xi h})-2n(V_{\overline\xi \xi \varphi}) -3 n(V_{h\bar\xi \omega})\0\\
&&+ 2n(\langle b\varphi\rangle) +  2n(\langle b\, h\rangle)+2n(\langle \overline \xi \xi\rangle) + 2n(\langle\overline\omega \omega\rangle)\label{UVdivgamma2}
\ee
Replacing \eqref{toprel} into \eqref{IRdivgamma1} we get
\be
 &&r(\gamma)= 4 -2n(V_{EH})-2n(V_{(2)\varphi h})  -2 n(V_{\overline\xi \xi h}) -2n(V_{ \xi \bar \xi \varphi})-3 n(V_{h\bar\xi \omega}) \0\\
 &&+ 2n(\langle h\, h\rangle)+ 2n(\langle h\, b\rangle)+  2n(\langle \varphi\, b\rangle)+2 n(\langle \overline \xi \xi\rangle)+ 2n(\langle\overline\omega \omega\rangle)\label{IRdivgamma2}
 \ee
Next we have to fix the degree of subtractions $\delta(\gamma)$ and $\rho(\gamma)$ for the UV and the IR, respectively, which is sensible to set at 4. So,
\be 
\delta(\gamma)&=&d(\gamma) + b(\gamma) , \quad\quad  \delta(\gamma) =4\label{UVdeg}\\
\rho(\gamma)&=& r(\gamma) -c(\gamma) , \quad\quad  \rho(\gamma) =4\label{IRdeg}
\ee
In order to make sure that the relevant Feynman integral is convergent in accordance with the propositions and theorems  proved in \cite{Lowen76} for truncated amplitudes we have to show that both $b(\gamma)$ and $c(\gamma)$ are non-negative integers. It is evident that the task of proving these relations is enormously simplified if we choose the gauges $\alpha_0=\beta_0=0$. From now on in this section we shall make this choice, postponing to the next section the task of proving that physical quantities are not affected by this choice.

Let us start with $b(\gamma)$. Remark that since $\gamma$ is 1PI, for any $\langle \overline \xi \xi\rangle$ internal line there must be at least one $V_{\bar \xi \xi h}$ or one $V_{\overline\xi \xi \varphi}$. Therefore $n(V_{ \overline\xi \xi h})+n(V_{\overline\xi \xi \varphi})\geq n(\langle \overline \xi \xi\rangle)$. Moreover any internal line $\langle b\varphi\rangle$ can end only in a vertex $V_{b\varphi h}$. Same for an internal line   $\langle b\, h\rangle$. Therefore $V_{b\varphi h}\geq n(\langle b\varphi\rangle)+ n(\langle b\, h\rangle)$. As a consequence $b(\gamma)\geq 0$. Let us notice that $\langle \overline \omega \omega\rangle $ does not play any role in the set of 1PI graphs because there is no vertex involving $\overline \omega$.

As for $c(\gamma)$ it is easy to construct a counterexample with $c(\gamma)< 0$. It is enough  to consider a one-loop graph with at least one vertex $V_{h\bar\xi \omega}$ (with external line $\omega$). Therefore we cannot satisfy the IR Lowenstein condition. On the other hand we notice that in any 1PI graph $V_{h\bar\xi \omega}$ can only be a peripheral vertex because the propagator $\langle \omega \bar\omega\rangle$  does not have any vertex involving $\overline \omega$ to which it can be attached to. This suggests that we can consider the subsets of 1PI diagrams that do not have $\omega$ external legs. For this subset of diagrams we can drop the term $ -3 n(V_{h\bar\xi \omega})$ both in \eqref{UVdivgamma2} and \eqref{IRdivgamma2}.

After this restriction of diagrams the UV Lowenstein condition keeps holding.
As for the IR one we notice first that $c(\gamma)$ (excluding $V_{h\bar\xi \omega}$) contains propagators and vertices multiplied by a uniform  factor of 2. If this  factor were different for different propagators and vertices the following argument in general would not hold. Let us start from one loop diagrams constructed out of trivalent vertices and propagators. It is evident that the  number $n$ of vertices equals the number $n$ of propagators  (with $n$ external legs); therefore $c(\gamma)=0$. If we change some of the trivalent vertices with polivalent ones, then either the number of external legs increases or the diagram has more than one loop. Amplitudes with more than one  loop are obtained either by adding internal lines that connect more than trivalent peripheral vertices, as just mentioned; or by breaking a peripheral line with the insertion of one or more vertices (without increasing the number of external legs); or by adding completely internal vertices. In all cases the number of added propagators is greater than the number of added vertices. Therefore $c(\gamma)\geq 0$.  

This ia not all, given a graph $\gamma$ the complete set of conditions, beside $c(\gamma) \geq 0$ and $b(\gamma)\geq 0$,  include also conditions on the reduced diagrams $\overline \gamma= \gamma/ \lambda_1,\ldots, \lambda_n$, where $\lambda_1, \ldots, \lambda_n$ are 1PI subdiagrams. Reduced subdiagrams are constructed by contracting mutually disjoint  1PI subdiagrams $\lambda_1,\ldots,\lambda_n$, each to a point which becomes a vertex   
$V(\lambda_i) $ and to which the unit polynomial of momenta is assigned. The UR and IR degree of divergence of reduced dagrams are related to those of $\gamma$ by
\be
&&d(\gamma)= d(\gamma/ \lambda_1,\ldots, \lambda_n) + \sum_{i=1}^n d(\lambda_i) \label{dbargamma}\\
&&r(\gamma)= r(\gamma/ \lambda_1,\ldots, \lambda_n) + \sum_{i=1}^n r(\lambda_i) \label{rbargamma}
\ee
It is evident from the previous discussion that for any 1PI subdiagram $\lambda_i$ without $\omega$ external legs we have as well $c(\lambda_i\geq 0$ and $b(\lambda_i)\geq 0$.

The additional conditions to be satisfied for reduced diagrams are:
\be
&&\delta(\gamma)\geq  d(\gamma/ \lambda_1,\ldots, \lambda_n)+ \sum_{i=1}^n \delta (\lambda_i) ,\quad\quad \delta (\lambda_i)=4\label{addcond1}\\
&&\rho(\gamma)\leq  r(\gamma/ \lambda_1,\ldots, \lambda_n)+ \sum_{i=1}^n \rho (\lambda_i), \quad\quad \rho (\lambda_i)=4 \label{addcond2}\\
&& \rho(\gamma)\leq\delta(\gamma) +1\label{addcond3}
\ee
For instance, extract $d(\overline \gamma)$ from \eqref{dbargamma} and replace it into \eqref{addcond1}. Then use \eqref{UVdeg} and the analogs for each $\lambda_i$: $4= d(\lambda_i) + b(\lambda_i$) , summed over $i$. One gets
\be
4\geq 4-b(\gamma) - \sum_{i=n} ^n b(\lambda_i) \label{addcond1ver}
\ee
which is true because all the $b$'s are nonnegative.
Similarly
\be
4\leq  4+c(\gamma)+  \sum_{i=n} ^n c(\lambda_i) \label{addcondver}
\ee
True, because all the $c$'s are nonnegative.

The above guarantees that the integral in \eqref{hatJgamma} is absolutely convergent and that the Green functions are well-defined tempered distributions. At higher loop level other conditions need to be satisfied, but in this paper we limit ourselves mostly to one-loop considerations. Therefore the above conditions are sufficient.

\subsection{ST identity}

As pointed out previously one cannot disentangle conformal and diffeomorphism   symmetry and quantize them separately. 
This is so because the diffeomorphisms and conformal transformations form a semidirect product group, which we have called $\cal G$; it includes an action of diffeomorphisms over the conformal transformation parameter represented by $\delta_\xi \omega = \xi\! \cdot \!\partial \omega$. 
It is clear that we have to construct the relevant ST identity for such overall symmetry $\cal G$. We introduce the familiar BRST transformation symbol $\mathsf s$ setting
\be
\mathsf s= \delta_\xi + \delta_\omega, \quad\quad {\mathsf s}^2=0\label{ess}
\ee
For instance ${\mathsf s} h_{\mu\nu} = \xi^\lambda \partial_\lambda h_{\mu\nu} + \partial_\mu \xi ^\lambda g_{\lambda \nu} + \partial_\nu \xi ^\lambda g_{\mu\lambda}+ 2\omega \, h_{\mu\nu},$ etc.

Next, in order to deal with the various non-linear transformations, we introduce the action for external auxiliary fields, 
\be
S_{ext} = \int \left( K^{\mu\nu} {\mathsf s}h_{\mu\nu} +L_\alpha \,{\mathsf s} \xi^\alpha +K\, {\mathsf s}\varphi+ L \,{\mathsf s} \omega)\right) \label{Sext}
\ee
so that our total classical action is 
\be
S_{tot}= S+S_{ext}. \label{Stot}
\ee

At this stage it is useful to summarize ghost numbers and dimensions of all the fields we have introduced so far.
\begin{table}[h]
\begin{tabular}{||l|l|l|l|l|l|l|l|l|l|l|l|l||}
\hline\hline \quad\quad & $\quad h_{\mu\nu}$\quad &\quad $\varphi$ \quad&\quad $\xi^\alpha$\quad &\quad $\omega$\quad &\quad $K_{\mu\nu}$\quad &\quad K\quad &\quad $L_\alpha$\quad &\quad L\quad &\quad $\overline \xi_\alpha$ \quad& \quad$\overline \omega$\quad &\quad $b_\alpha$\quad & \quad b\quad\\ \hline
gh\#&\quad 0\quad&\quad 0\quad &\quad 1\quad & \quad 1 \quad & \quad -1\quad & \quad -1 \quad& \quad -2\quad & \quad -2 \quad &\quad -1 \quad & \quad -1 \quad &\quad 0 \quad& \quad 0\quad \\ \hline
dim&\quad 0\quad&\quad 0\quad &\quad -1\quad & \quad 0 \quad & \quad 4\quad & \quad 4 \quad& \quad 5\quad & \quad 4 \quad &\quad 2 \quad & \quad 2 \quad &\quad 2 \quad& \quad 2\quad 
\\\hline\hline
\end{tabular}
\caption{Ghost number (gh$\#$) and physical dimension (dim)}
\label{tab:ghost}
\end{table}

Now, the ST identity for the vertex functional $\Gamma$ takes the form
\be
{\mathbb S} (\Gamma) = \int \left( \frac {\delta \Gamma}{\delta K^{\mu\nu} }\frac  {\delta \Gamma}{\delta h_{\mu\nu}}+  \frac {\delta \Gamma}{\delta L_\alpha}\frac  {\delta \Gamma}
{\delta \xi^\alpha}  + \frac {\delta \Gamma}{\delta K} \frac {\delta \Gamma}{\delta \varphi}+ \frac {\delta \Gamma}{\delta L} \frac {\delta \Gamma}{\delta \omega}-b_\alpha \frac {\delta \Gamma }{\delta \overline\xi^\alpha} - b \frac {\delta\Gamma}{\delta\overline\omega}\right)=0 \label{STGamma}
\ee
This equation expresses the invariance of $\Gamma$ under $\cal G$ transformations. It is certainly true  when $\Gamma$ is identical to the total classical action $S_{tot}$. The challenge of renormalization is to prove that it remains true at any order of approximation of the expansion
\be
\Gamma=\sum_{n=0}^\infty {\slashed h}^n \Gamma^{(n)}, \quad\quad {\rm where} \quad\quad \Gamma^{(0)}= S_{tot} .  \label{Gammaexp}
\ee
In this context an important role is played by the linearized version of ${\mathbb S}$
\be
{\EuScript S}_{\Gamma} &=&  \int \left( \frac {\delta  \Gamma}{\delta  K^{\mu\nu} }\frac  {\delta}{\delta h_{\mu\nu}}+\frac  {\delta \Gamma}{\delta h_{\mu\nu}}  \frac {\delta}{\delta K^{\mu\nu} }+  \frac {\delta \Gamma }{\delta L_\alpha}\frac  {\delta}
{\delta \xi^\alpha}  +\frac  {\delta \Gamma}
{\delta \xi^\alpha}\frac  {\delta}
{\delta L_\alpha} + \frac {\delta  \Gamma}{\delta\widehat K} \frac {\delta  }{\delta \varphi}+\frac {\delta \Gamma}{\delta \varphi} \frac {\delta}{\delta  K}+ \frac {\delta \Gamma}{\delta L} \frac {\delta }{\delta \omega}+\frac {\delta  \Gamma}{\delta \omega}\frac {\delta}{\delta L}\right.\0\\
&&\left.-b_\alpha \frac {\delta  }{\delta \overline\xi^\alpha} - b \frac {\delta }{\delta\overline\omega} \right)\label{STGammalinear}
\ee
This operator is nilpotent
\be
{\EuScript S}_{\Gamma} {\EuScript S}_{\Gamma} =0 \label{SGammanil}
\ee

Beside \eqref{STGamma} two gauge fixing conditions 
\be
\frac{\delta \Gamma}{\delta b^\mu} = \frac 1{\sqrt \kappa} \left( \partial^\nu h_{\mu\nu}-2 \partial_\mu \varphi\right) - {\alpha_0}  b_\mu \label{gfcond1}
\ee
and 
\be
\frac{\delta \Gamma}{\delta b} = -\square \varphi + \beta_0 b\label{gfcond2}
\ee
and two ghost equations 
\be
\frac{\delta \Gamma}{\delta \overline \xi^\mu} - \partial_\nu\left( \frac{\delta\Gamma}{\delta K_{\mu\nu}}-2 \eta^{\mu\nu} \frac {\delta\Gamma}{\delta K}\right) =0\label{ghosteq1}
\ee
and
\be
\frac{\delta \Gamma}{\delta \overline \omega} - \square \frac {\delta\Gamma}{\delta K} =0\label{ghosteq2}
\ee
must be satisfied order by order.

If we express $\Gamma$ as a function of $\hat K_{\mu\nu} = K_{\mu\nu}+\frac 12 \left( \partial_\mu \overline \xi_\nu+  \partial_\nu \overline \xi_\mu\right)$, then \eqref{ghosteq1} can be rewritten as
\be
\frac{\delta \Gamma}{\delta \overline \xi^\mu}+ 2\partial_\mu  \frac {\delta\Gamma}{\delta K}=0
\label{ghosteq1'}
\ee
and if we, moreover, express $\Gamma$ in terms of $\hat K=K + 2\partial_\nu\overline \xi^\nu - \square\, \overline \omega$ , the two ghost equations can be rewritten simply as
\be
\frac{\delta \Gamma}{\delta \overline \xi^\mu}=0 \label{ghosteqfin1}
\ee
and
\be
\frac{\delta \Gamma}{\delta \overline \omega} =0,\label{ghosteqfin2}
\ee
respectively. With the passage from $K_{\mu\nu}, K$ to $\widehat K_{\mu\nu}, \widehat K$, the last two terms of \eqref{STGamma} disappear. In Appendix B it is proven that the gauge fixing conditions and the ghost equations are not renormalized by quantization. Therefore it is convenient to define a functional $\widehat \Gamma$ independent of $b_\alpha$ and $b$:
\be
\widehat \Gamma(h_{\mu\nu},\varphi,\xi^\alpha,\omega,\widehat K_{\mu\nu}, \widehat K, L_\alpha, L)= \Gamma(h_{\mu\nu},\varphi,\xi^\alpha,\omega,b_\alpha, b,\overline \xi^\alpha, \overline\omega, K_{\mu\nu}, K, L_\alpha, L) -S_{g.f.}^{(diff,c)} -S_{g.f.}^{(c)}\label{Gammaredef}
\ee
see \eqref{gfdiffconf}, \eqref{gfconf}. With these redefinitions the  ST identity  \eqref{STGamma} becomes
\be
{\mathbb S} (\widehat\Gamma) = \int \left( \frac {\delta \widehat\Gamma}{\delta\widehat K^{\mu\nu} }\frac  {\delta \widehat\Gamma}{\delta h_{\mu\nu}}+  \frac {\delta \widehat\Gamma}{\delta L_\alpha}\frac  {\delta \widehat\Gamma}
{\delta \xi^\alpha}  + \frac {\delta \widehat\Gamma}{\delta\widehat K} \frac {\delta \widehat\Gamma}{\delta \varphi}+ \frac {\delta \widehat\Gamma}{\delta L} \frac {\delta \widehat\Gamma}{\delta \omega}\right)=0.\label{SThatGamma}
\ee
This can be expressed in terms of the linear functional operator ({\it the linearized ST operator})
\be
{\EuScript S}_{\widehat\Gamma} =  \int \left( \frac {\delta \widehat\Gamma}{\delta\widehat K^{\mu\nu} }\frac  {\delta}{\delta h_{\mu\nu}}+\frac  {\delta \widehat\Gamma}{\delta h_{\mu\nu}}  \frac {\delta}{\delta\widehat K^{\mu\nu} }+  \frac {\delta\widehat\Gamma }{\delta L_\alpha}\frac  {\delta}
{\delta \xi^\alpha}  +\frac  {\delta \widehat\Gamma}
{\delta \xi^\alpha}\frac  {\delta}
{\delta L_\alpha} + \frac {\delta \widehat\Gamma}{\delta\widehat K} \frac {\delta  }{\delta \varphi}+\frac {\delta \widehat\Gamma}{\delta \varphi} \frac {\delta}{\delta\widehat K}+ \frac {\delta \widehat\Gamma}{\delta L} \frac {\delta }{\delta \omega}+\frac {\delta \widehat\Gamma}{\delta \omega}\frac {\delta}{\delta L}\right) \label{STlinearop}
\ee
as 
\be
{\EuScript S}_{\widehat\Gamma}(\widehat \Gamma)= 2\, {\mathbb S} (\widehat\Gamma)=0 \label{STlinear}
\ee
It is clear that this passage from $\Gamma$ to $\widehat \Gamma$ and from \eqref{STGamma} to \eqref{SThatGamma} is possible only if \eqref{gfcond1} and \eqref{gfcond2}, as well as \eqref{ghosteqfin1} and \eqref{ghosteqfin2} are preserved during the renormalization process. This issue is treated in Appendix B.

Here we continue examining the consequences of  \eqref{STlinear}. 
One can also prove that if \eqref{STlinear} is true then
\be
\left({\EuScript S}_{\widehat\Gamma}\right)^2=0\label{STlinearopsquare}
\ee
which is the analog of \eqref{STGammalinear}.

It is evident that, if we apply ${\EuScript S}_{\widehat\Gamma}$ to $h_{\mu\nu}$ or $\varphi$ this operation gives back the BRST transforms of these fields. Same for $\xi^\alpha$ and $\omega$. But what if we apply it to the external fields? The first term on the RHS of \eqref{SThatGamma} tells us that, if we replace differentiably $h_{\mu\nu}$ wherever in $\widehat \Gamma$ with its BRST transform we get 0. But the same terms can be also read in another sense: if we replace differentiably $\widehat K_{\mu\nu} $ with the quantity expressed by $\frac  {\delta \widehat\Gamma}{\delta h_{\mu\nu}} $, we get 0 as well. That is, we can interpret the latter quantity as an extension of the BRST  transformation. This together with \eqref{STlinearopsquare} means that we can extend the BRST transformations \eqref{sBRST}, \eqref{sBRST2} and \eqref{sbaromega}, to include also the external fields by defining a new operator $\mathfrak b$ that extends $\mathsf s$ as follows
\be
{\mathfrak b}\, \phi\equiv {\EuScript S}_{\widehat S}\,( \phi)\label{mathsfb}
\ee
for any field $\phi$. Here $\widehat S$ is the classical action $S_{tot}$ minus the gauge fixing terms \eqref{gfdiffconf} and \eqref{gfconf}. Thus $\mathfrak b$ is the same as $\mathsf s$ for $h_{\mu\nu}, \varphi, \xi^\alpha, \omega,b_\alpha, b, \overline \xi_\alpha, \overline \omega$. In addition we have the extended BRST transformation of $K_{\mu\nu}, K, L_\alpha, L$ given by the formula \eqref{mathsfb}:
\be
{\mathfrak b}\,  K^{\mu\nu}&=& \frac {\delta S_0} {\delta h_{\mu\nu}}-\partial_\lambda \xi^\lambda  K^{\mu\nu}+\partial_\lambda \xi^\mu  K^{\lambda\nu}+\partial_\lambda \xi^\nu  K^{\mu\lambda}- 2 \omega \,  K^{\mu\nu}\0\\
{\mathfrak b}\, K &=&\frac {\delta S_0} {\delta\varphi} - \partial_\lambda \left( \xi^\lambda K \right)\0\\
{\mathfrak b}\, L_\alpha &=&\partial_\alpha \xi^\lambda L_\lambda -\partial_\lambda \left(\xi ^\lambda L_\alpha\right)\0\\
{\mathfrak b}\, L &=&\partial_\lambda \left(\xi ^\lambda L\right)\label{btransf}
\ee
where $S_0$ is given in eq.\eqref{S0}.

Before continuing we need another formal relation valid for any functional  $\digamma$:
\be
{\EuScript S}_{\digamma} {\mathbb S}({\digamma})=0\label{digamma}
\ee

The action of ${\mathbb S}$ on $\Gamma$ produces terms  of dimension 4 and ghost number 1,
That is, from \eqref{STGamma} and \eqref{Gammaexp}
\be
{\mathbb S}(\Gamma)= \sum_{n=1}^\infty \, \hslash ^n \Delta^{(n)}\label{Deltan}
\ee
and the quantum action principle, \cite{piguetsorella}, we deduce that $\Delta^{(n)}$ are integrated polynomials of the fields and their derivatives of dimension 4 and ghost number 1. As a consequence of \eqref{digamma} the quantum corrections $\Delta^{(n)}$ must satisfy.
\be
{\mathfrak b} \Delta^{(n)}=0 \label{bDeltan}
\ee
and since $\mathfrak b$ is nilpotent this equation defines a cohomology. A fundamental step of the renormalzation program is the determination of this cohomology. 

The first question concerning such cohomology is whether the external fields play a role in it. This problem is analysed in Appendix C and the answer is that there is indeed a possible $\Delta$ involving the external field $L$
\be
\Delta(L) = \int  \xi\!\cdot\! \partial\xi \!\cdot\! \partial \omega\, L\label{DeltaLcob}
\ee
which satisfies the condition ${\mathfrak b} \Delta(L)=0$. But this is a coboundary which can be reabsorbed by subtracting a counterterm
\be
C(L)=\int \xi\!\cdot\! \partial\xi \!\cdot\! \varphi\,L \label{C(L)count}
\ee
for ${\mathfrak b} C(L)= \Delta(L)$. 

This is the only coboundary in the cohomology involving external fields, while there are no cocycles. Furthermore, the subtraction of the counterterm $C(L)$ does not affect neither the gauge fixing conditions (\ref{gfdiffconf}, \ref{gfconf}) nor the ghost equations (\ref{ghosteqfin1},\ref{ghosteqfin2}).

\subsection{Gauge independence}

A crucial question for the validity of the renormalization process concerns the independence of the two gauge-fixing conditions we have chosen, \eqref{gfdiffconf} and \eqref{gfconf}. To prove this independence there exists a well known formal trick. It consists in enlarging the BRST algebra by giving the gauge constant $\alpha_0$ and $\beta_0$ the status of variables in the algebra, with anticommuting partners $\chi_{\alpha_0}$ and $\chi_{\beta_0}$, with transformations:
\be
&&{\mathsf s} \alpha_0= \chi_{\alpha_0} , \quad\quad   {\mathsf s} \chi_{\alpha_0}=0\label{addalpha0}\\
&&{\mathsf s} \beta_0= \chi_{\beta_0} , \quad\quad   {\mathsf s} \chi_{\beta_0}=0\label{addalpha0}
\ee
This amounts to adding  \eqref{gfdiffconf} and \eqref{gfconf} the terms
\be
- \frac {\chi_{\alpha_0}}2 \int \eta_{\mu\nu} b^\mu b^\nu= {\mathsf s} \frac {\chi_{\alpha_0}}2 \int \eta_{\mu\nu} \overline \xi^\mu b^\nu\quad\quad {\rm and} \quad\quad\frac {\chi_{\beta_0}}2\int b^2= -{\mathsf s}\frac {\chi_{\beta_0}}2\int \overline \omega\,b, \label{gfadd}
\ee
respectively. Now we define a new ST operator and identity
\be
{\mathbb S}^{(new)}(\Gamma) = {\mathbb S}^{(old)} ( \Gamma)+ \chi_{\alpha_0} \frac {\delta  \Gamma} {\delta \alpha_0} +\chi_{\beta_0} \frac {\delta   \Gamma} {\delta \beta_0}=0 \label{newST}
\ee
Differentiating both sides with respect to $\chi_{\alpha_0}$ we find
\be
 \frac {\delta \Gamma} {\delta {\alpha_0}}= {\EuScript S}^{(new)}_{\widehat \Gamma}\left( \frac {\delta  \Gamma} {\delta {\alpha_0}}\right)\label{alpha0ind}
 \ee
Since ${\EuScript S}^{(new)}_{\widehat \Gamma}$ is the linearized ST operator  representing the extended $\mathfrak b$ BRST operator, \eqref{alpha0ind} tells us that any dependence on $\alpha_0$ is a BRST transform. Since physical quantities are BRST invariant ones of dimension 4 and ghost number 0, it follows that the RHS of \eqref{alpha0ind} is a nonphysical quantity. A similar result can be obtained by differentiating \eqref{newST} with respect to $\chi_{\beta_0}$, which tells us that any dependence on $\beta_0$ is a BRST transform  too.

We must hasten to specify that the compliance of our derivation with the convergence conditions for 1PI diagrams has been proven above with the gauge choice $\alpha_0=\beta_0=0$. This does not mean that in other gauges the conditions are not satisfied. We can reasonably suppose that the convergence conditions are satisfied, if not for all, at least for a dense enough set of amplitudes to justify the use of the functional differential calculus in the previous formulas, and in particular for \eqref{alpha0ind}. However an explict proof is lacking and would be most desirable.

\section{Weyl invariants and (trivial) Weyl cocycles} 
\label{s:weylcohomology}

In this section we carry out the analysis of the $\mathfrak b -$cohomology introduced in the previous one. We consider a Weyl invariant theory involving the graviton and the dilaton, such as \eqref{S0} for instance, and wish to determine the possible Weyl cocycles that may emerge from it. The differential space we are interested in is constructed out of the metric $g_{\mu\nu} $ and the dilaton $\varphi$, together with the diffeomorphism parameter $\xi^\mu$ and Weyl parameter $\omega$. The latter two are promoted to anticommuting fields. It is known that in 4d there are no consistent anomalies due only to the diffeomorphisms. Therefore we will assume general diffeomorphism symmetry and search for Weyl cocycles that respect diffeomorphism invariance\footnote{A cocycle of the diffeomorphism+Weyl cohomology, which is not reducible to a diffeomorphism-covariant form, is not known, and is  excluded here.}. Therefore the cocycles will be expressions of physical  dimension 4 made of the Riemann tensor, of the Ricci tensor and scalar, their covariant derivatives, as well as of covariant derivatives of the dilaton with coefficients that may be functionally differentiable functions of the dilaton field. The cocycles we are interested in are of zeroth (invariants) and first order (anomalies) in $\omega$ (which is dimensionless). In order to study invariants and anomalies in the case in which the dilaton plays an active role, it is more convenient to express the differential space in terms of  conformal invariant objects instead of the ordinary ones. That is, instead of $R^\mu{}_{\nu\lambda\rho}, R_{\mu\nu}$ and $R$ we shall use $\widetilde R^\mu{}_{\nu\lambda\rho},\widetilde R_{\mu\nu}$ and $\widetilde{\widetilde R}= e^{2\varphi} \widetilde R$, and instead of the ordinary covariant derivative $D_\mu$ the Weyl covariant one $\widetilde D_\mu$, as well as the conformal invariant metric $\widetilde g_{\mu\nu} = e^{-2\varphi} g_{\mu\nu}$. At any stage it is possible to shift back to the ordinary notation.

\subsection{Weyl cocycles with $\varphi$}

To save space we shall use a shorthand notation for a basis of independent 0-cycles $C_i(\widetilde g,\varphi)$ of the differential space as follows:
\be
C_i(g,\varphi)=\int d^4x \sqrt{\widetilde g} f_i(\varphi) {\mathfrak c}_i(\widetilde g,\varphi), \quad \quad i=1,\dots, 11.\label{Ci}
\ee
where $f_i({\varphi})$ are generic differentiable functions of $\varphi$, and ${\mathfrak c}_i$ are given in the following list numbered from 1 to 7,
\be 
\widetilde \square \varphi\widetilde \square \varphi,\,\, \widetilde \square \varphi  D_\mu \varphi D^\mu\varphi, \,\, D_\mu \varphi D^\mu\varphi D_\nu \varphi D^\nu\varphi, \,\,D^\mu \varphi D^\nu\varphi \widetilde D^\mu D^\nu \varphi,\,\,\widetilde D^\mu D^\nu \varphi\widetilde D_\mu D_\nu \varphi, \,\, \widetilde{\widetilde R}D_\mu \varphi D^\mu\varphi, \widetilde{\widetilde R}\, \widetilde \square \varphi,\label{list1}
\ee
and from 13 to 15,
\be 
\widetilde{\widetilde R}^2, \quad \widetilde R_{\mu\nu} \widetilde R^{\mu\nu}, \quad
\widetilde R_{\mu\nu\lambda\rho} \widetilde R^{\mu\nu\lambda\rho}.\label{list2}
\ee
{ In this subsection repeated indices are understood to be saturated with the metric $\widetilde g_{\mu\nu}$. The symbol $\widetilde \square$ means $ \widetilde g^{\mu\nu}\widetilde D_\mu \partial_\nu=  \widetilde g^{\mu\nu}\left(\partial_\mu \partial_\nu - \widetilde \Gamma_{\mu\nu}^\lambda \partial_\lambda\right) $. }

We next apply the nilpotent functional operator $\delta_\omega$ \footnote{The BRST operator $\mathsf s$ reduces to $\delta_\omega$ because $\delta_\xi$ acts trivially on covariant expressions.} to the first group and, by integrating by parts, we put the results in the form (undifferentiated $\omega$):
\be
\delta_\omega \sum_{i=1}^7 C_i(\widetilde g,\varphi)=\int d^4x \sqrt{\widetilde g}\, \omega \sum_{i=1}^7{\mathfrak C}_i(f_i, \widetilde g, \varphi)\label{mathCi}
\ee
${\mathfrak C}_i$ contain, beside terms like the ones in \eqref{list1}, also functional derivatives of $f_i$ with respect to $\varphi$. The invariants are obtained by equating to 0 the RHS of \eqref{mathCi}, which  imposes linear conditions on $f_i$ and derivatives thereof. For instance the term containing 
$D_\mu \varphi D^\mu\varphi D_\nu \varphi D^\nu\varphi, $ yields the condition
\be
\frac {\delta^2 f_2}{\delta\varphi ^2}- 3 \frac {\delta f_3}{\delta\varphi}+ \frac {\delta^2 f_4}{\delta\varphi ^2}+ \frac {\delta^3 f_3}{\delta\varphi ^3}=0
\ee
There are 13 such conditions, and it is not hard to see that the only solution is $f_i=0$ for all $i=1,\ldots, 7$.

The analysis is simpler for the three terms $C_{13},C_{14}$ and $C_{15}$. The invariance conditions imposes that $f_{13},f_{14}$ and $f_{15} $ are constant. With this specification they represent the three conformal invariants  of the theory. As it has already been remarked a linear combination of the three is the Euler or Gauss-Bonnet density, which is a total derivative and thus does not count in the framework of perturbative field theory.  The two surviving cycles, written in the standard untilded notation, can be chosen to be the actions $S_C$ (or $S'_C$) and $S_Q$, see eqs.(\ref{Weyltensoraction},\ref{Sc'W},\ref{intQ2}).

In the action $S_0$ there is also an invariant of dimension 2, proportional to $\widetilde{\widetilde R}$. It is the only one such term.

The computation of 1-cocycles is understandably more complicated. We start from a list of 1-cycles
\be
\Delta^{(1)}_i(\omega,f_i, \widetilde g,\varphi)= \int d^4x \sqrt{\widetilde g}\, \omega\, f_i(\varphi) {\mathfrak D}_i (\widetilde g,\varphi)\label{Deltai}
\ee
where the ${\mathfrak D}_i (\widetilde g,\varphi)$ are the same as the ${\mathfrak c}_i$ of the list \eqref{list1}, plus the following ones numbered from 8 till 12
\be 
\widetilde \square^2 \varphi, \quad D^\mu\varphi \widetilde D_\mu\widetilde \square \varphi,\quad 
D_\mu \widetilde{\widetilde R}\,  D^\mu \varphi, \quad \widetilde R_{\mu\nu} D^\mu\varphi D^\nu \varphi,\quad  \widetilde R_{\mu\nu}\widetilde D^\mu D^\nu\varphi\label{list3}
\ee
When we apply $\delta_\omega$ to the 1-cycles $ \Delta^{(1)}_i(\omega,f_i,\widetilde g,\varphi)$ we obtain 2-cycles, $ \Delta^{(2)}_i(\omega,f_i,\widetilde g,\varphi)$, i.e. bilinears in $\omega$. For instance
\be
\delta_\omega \Delta^{(1)}_1(\omega,f_i,\widetilde g,\varphi)= -2\int d^4x \sqrt{\widetilde g}\, \omega\, \widetilde \square\omega \,\widetilde \square\varphi\label{C1}
\ee

In general the $ \Delta^{(2)}_i(\omega,f_i,\widetilde g,\varphi)$  can be expressed  in terms of the product of two $\omega$'s separated by an order $n$ differential operator, multiplied by an $f_i$ or functional derivative thereof, multiplied by a dimension $4-n$ polynomial made of spacetime derivatives  of  $\varphi$ and possibly of the conformal invariant Ricci scalar and tensor and their derivatives. The products of such polynomials times the $\omega$ bilinears, can be organized as sums of 15 distinct terms, which, however, are not all independent. This is due to the nilpotence of $\omega$. For instance we have
\be
0= \int d^4x \sqrt{\widetilde g} \,f(\varphi)\, \widetilde \square \omega \,  \widetilde \square \omega=
 \int d^4x \sqrt{\widetilde g} \,\left(f'' \omega \widetilde \square \omega D^\mu\varphi D_\mu\varphi + 2 f'\, \omega  \widetilde D^\mu \widetilde \square \omega D_\mu\varphi + f\,  \omega \widetilde \square^2 \omega\right),\label{0ident}
 \ee
which linearly connects three such terms (primes denote functional derivatives with respect to $\varphi$). There  are 7 independent relations like \eqref{0ident}. Using these we can express all the $ \Delta^{(2)}_i(\omega,f_i,\widetilde g,\varphi)$ as linear sums of 8 such terms with coefficients consisting of linear sums of $f_i$ and functional derivatives thereof. To obtain 1-cocycles we have to impose the vanishing of the coefficients of such 8 terms. Since the system is linear, supposing the rank is maximal, that is 7, it determines 7 independent solutions up to a multiplicative factor. This means that we have 7 independent 1-cocycles. They are trivial because they necessarily coincide with the cocycles we obtain by applying $\delta_\omega$ to the seven independent 0-cycles of the list \eqref{list1}. So much for the list numbered from 1 to 12 built from the list \eqref{list1} and \eqref{list3}.

This of course is not the end of our analysis because we have to consider also the  $ \Delta^{(1)}_i(\omega,f_i,\widetilde g,\varphi)$ obtained through formula \eqref{Deltai} from the list \eqref{list2} with the addition of the term $\widetilde \square{\widetilde{\widetilde R}}$, numbered 16. Applying $\delta_\omega$ to the 14-th member of this list, for instance, we get
\be
\delta_\omega \int d^4x \sqrt{\widetilde g}\,\omega\,f_{14}(\varphi) \widetilde R_{\mu\nu} \widetilde R^{\mu\nu} = 0\label{Delta14}
\ee
The same for the other three. Therefore these are 1-cocycles and we have to verify whether they are true cocycles or coboundaries. We remark that
\be
\delta_\omega \int d^4x \sqrt{\widetilde g}\,g(\varphi) \widetilde R_{\mu\nu} \widetilde R^{\mu\nu} = \int d^4x \sqrt{\widetilde g}\,g'(\varphi) \,\omega \widetilde R_{\mu\nu} \widetilde R^{\mu\nu}\label{deltaDelta14}
\ee
The same for the other three cases. For \eqref{Delta14} it is enough to choose a $g(\varphi)$ such that $g'= f_{14}$.  Therefore we have four coboundaries (trivial cocycles). 

There are no true anomalies in terms of the fields $\widetilde g_{\mu\nu}, \varphi$ and their conformal covariant derivatives alone. We have to consider also another possibility: cocycles of $\delta_\omega$ expressed only in terms of the metric without any role of $\varphi$.

\subsection{Weyl cocycles without $\varphi$}

Any trace anomaly of this type can be written in the form
\be
{\cal A}_\omega[g,f] = \int d^4x \,\sqrt{g}\,\omega\, {\cal F}[g,f] \label{traceanomaly}
 \ee
 where $g=\{g_{\mu\nu}\}$ is the metric, and $f$ represents any other field. ${\cal A}_\omega[g,f] $ satisfies the consistency condition
 \be
 \delta_\omega\, {\cal A}_\omega[g,f] =0\label{deltaomegacc}
 \ee
 All the solutions of this cohomology problem with $ {\cal F}[g,f]={\cal F}[g] $, depending only on the metric and no other fields, have been found over the years. They are those  where the density $\cal F$ takes the form of the quadratic Weyl density
\be
\EW^2=R_{\mu\nu\lambda\rho} R^{\mu\nu\lambda\rho}-2 R_{\mu\nu}R^{\mu\nu} +\frac 13 R^2,\label{quadweyl}
\ee
the Gauss-Bonnet (or Euler) density,
\be
E=R_{\mu\nu\lambda\rho} R^{\mu\nu\lambda\rho}-4 R_{\mu\nu}R^{\mu\nu} +R^2,\label{gausbonnet}
\ee
the Pontryagin density,
\be
P=\frac 12\left(\varepsilon^{\mu\nu\mu'\nu'}R_{\mu\nu\lambda\rho}R_{\mu'\nu'}{}^{\lambda\rho}\right). \label{pontryagin}
\ee 
and
\be
{\cal F}[g]=\square R\label{squareR}
\ee
which is known to be trivial, while the previous ones are not. 

Concerning odd parity anomalies, we have constructed the theory $\cal T$ so as to get rid of them (and they are not modified by introducing $\varphi$), see also the Comment at the end of section 8. For the even parity trace anomalies the story is different. Let us recall first that they do not obstruct the existence of propagators, therefore they are not dangerous from the point of view of quantization. But the even parity trace anomalies have the same sign in both chiral sectors and the coefficients in front of them are so random that it is impossible to cancel them adding up different species, except perhaps in very exotic models. In order to ensure the survival of conformal invariance while preserving locality we need counterterms. As a matter of fact they are all trivial when a dilaton is present in the model. This is obvious for \eqref{quadweyl} since it can be rewritten in the same form by means of tilded quantities and $\sqrt{\widetilde g}$ in \eqref{traceanomaly} due to the conformal property of the Weyl tensor.  It is less obvious for \eqref{gausbonnet}. For it we need a Wess-Zumino term.

\subsection{Wess-Zumino terms}

As pointed out above any cocycle  ${\cal A}_\omega$ of $\delta_\omega$ satisfies the Wess-Zumino consistency condition
\eqref{deltaomegacc},  which expresses simply the fact that two subsequent Weyl transformations made in opposite order yield the same result. This is in fact an integrability condition. It means that, with the help of an auxiliary field $\sigma$, which transform as $\delta_\omega \sigma=-\omega$, we can construct a local functional ${\cal F}_{WZ}[\sigma,g,f]$, such that
 \be
 \delta_\omega  {\cal F}_{wz}[\sigma,g,f]= - {\cal A}_\omega[g,f] \label{WZtrace}
 \ee
It can be explicitly constructed as follows
\be
 {\cal F}_{wz} [\sigma,g,f] = \int_0^1 dt\, \int d^4x\,\sqrt{g(t)} F[g(t),f(t)]\,\sigma \label{WWZ}
 \ee
 where 
  \be
 g_{\mu\nu} (t) = e^{2\sigma t} g_{\mu\nu},\quad\quad {\rm so \,\,\,that}\quad\quad  \delta_\omega g_{\mu\nu}(t)= 2(1-t)\,\omega\,g_{\mu\nu}(t), \label{interpolatingmetric}
 \ee
 and 
 \be
 f(t) = e^{-y\, t\, \sigma} f, \quad\quad \delta_\omega f(t) = -y(1-t)\omega f(t)\label{f(t)}
 \ee
 where $y=0$ for a gauge field, $y=1$ for a scalar field. 

It can be easily proved that \eqref{WWZ} satisfies \eqref{WZtrace}.

Now, for any trace anomaly the strategy consists in adding to the effective action the corresponding WZ term and setting $\sigma=-\varphi$. In other words WZ terms are analogous of the counterterms of section 6.1. They guarantee that at one loop the theory is anomaly free, and conformal invariance is recovered. The price one has to pay, here as well as in subsection 6.1, is the addition of a number of WZ terms, which, from the perturbative quantization point of view, are renormalizable interaction terms. It must be recalled that such WZ terms cannot appear in the classical theory (0-th order level) because they would break conformal invariance. The coefficients in front of the WZ terms will be a series whose lowest term is proportional to $\hslash$. 

This result together with the conclusion of subsection 6.1 amounts to the statement that there are no conformal anomalies in conformal field theories involving a metric and a dilaton.

\section{How to kill quartic derivatives terms}

Once the cohomology of $\mathfrak b$ is known the renormalization program can proceed. One must calculate the true conformal anomalies of the theory and subtract the corresponding counterterms in order to cancel them at one-loop. One first remark is that all the counterterms introduced above in order to trivialize the anomalies are all interaction terms and do not modify the structure of the kinetic terms  of the theory. As already noted, these counterterms cannot be introduced at the zeroth order of perturbation because they are not conformal invariant. Therefore they appear only from one-loop order on.
 The next steps in the renormalization would be to define the normalization conditions, write down the renormalization group equation and so on. We would also wish to extend the previous constructions by adding to $S_0$ fermionic and bosonic matter. Altogether this does not seem to imply insurmountable difficulties and the renormalization program (with the restriction introduce above on $\Gamma$) is likely to be viable. This part will be continued elsewhere. Now, we would like to focus on the other big problem, the existence of quartic poles in some of the propagators, which are a threat to unitarity. They come from the terms $S_C$ (or $S_C$) and $S_Q$, eqs.(\ref{Weyltensoraction},\ref{Sc'W},\ref{intQ2}). These terms can be easily written in terms of the tilded (conformal invariant) quantities $\widetilde R_{\mu\nu\lambda\rho}, \widetilde R_{\mu\nu}, \widetilde{\widetilde R}$ and $\widetilde g_{\mu\nu}$. $S_C$ or $S'_C$ can be rewritten in the same form by replacing the original symbols with the just mentioned tilded ones. As for $S_Q$ it can be rewritten
as
\be 
S'_Q=\frac 1{6\gamma}  \int d^4x \sqrt{\widetilde g}\,\widetilde{\widetilde R}^2\label{intQ2'}
\ee
Let us focus on $S_C$. We have seen above that its density is a superposition of the three terms in \eqref{list2}. The latter are the densities of three coboundaries given by the formula \eqref{Deltai}, with i=13,14,15, and $f_i=1$.  The resulting coboundary  is canceled by the counterterm
\be
C(W)=\int d^4x\,\sqrt{\widetilde g}\, \varphi \left(\widetilde R_{\mu\nu\lambda\rho} \widetilde R^{\mu\nu\lambda\rho}-2\widetilde R_{\mu\nu}\widetilde R^{\mu\nu} +\frac 13\widetilde{\widetilde R}^2\right)\label{quadweylcounter}
\ee 
This counterterm $C(W)$ multiplied by $\hslash$ must be subtracted from $\Gamma$ at the first loop order (not from the classical action because it is not covariant). This counterterm guarantees conformal invariance at the one-loop order. On the other hand if $\varphi$ takes a constant value, $C(W)$ matches exactly the form of $S_C$ apart from an overall constant. Adjusting the value of $\varphi$ the two terms kill each other exactly. Therefore, for this particular value of $\varphi$ the quartic derivative terms coming from $S_C$ disappear and do not introduce physical ghosts in the theory.

It is clear that we can use the same argument also for the term $S_Q$. If the anomaly whose density is proportional to $\widetilde{\widetilde R}^2$ is present in the theory, as the above analysis suggests, we have to cancel it by the counterterm
\be
C(Q)=\int d^4x\,\sqrt{\widetilde g}\, \varphi \, \widetilde{\widetilde R}^2\label{Q2counter}
\ee 
By choosing a specific constant value for $\varphi$ we can suppress the term $S_Q$. Therefore all the quartic derivative terms in the theory disappear in this specific gauge of $\varphi$. If all the quartic terms disappear the way is open to demonstrate the absence of unphysical particles, i.e. to prove unitarity.  But the theory is conformal invariant and its action does not depend on the value of $\varphi$. Therefore this conclusion should be valid in general.

Two important specifications are in order. First, the constant value taken by $\varphi$ cannot in general satisfy the cancellation of both $S_C$ and $S_Q$. We need two dilaton fields. In \cite{bonora25} it  was shown how this can be achieved. On the other hand the necessity of more than one dilaton field comes also from the multiplicity of scales that are needed in order to describe a complete theory of the universe evolution. In a cosmological framework the gigantic jump from the electroweak scale to the tiny cosmological constant is not the only one. We may need other intermediate scales. They can be inserted in our scheme as follows. With reference to the beginning of section 4, in particular to eq.\eqref{Rscalar}, let us introduce two dilaton fields $\varphi_1$ and $\varphi_2$, which transform like $\varphi$ above under diffeomorphisms and conformal transformations, and define
\be
\widetilde R_{12} = R + 6\left(\partial \!\cdot\! S -S\cdot S\right), \label{Rscalar12}
\ee
where
\be
S_\mu=\epsilon\, \partial_\mu \varphi_1 +(1-\epsilon)\, \partial_\mu\varphi_2\label{Smu}
\ee
and $\epsilon$ is a real number. 
Next consider, as an example, the action
\be
S_{12}^{(c)} &=&-\frac 1{2\kappa} \int d^4x \sqrt{g}\, e^{-2\varphi_1} \left(\widetilde R_{12} + {\mathfrak c}\, e^{-2\,\varphi_1}\right)\0\\
&&+ \int d^4 x \,\sqrt{g} \left[ g^{\mu\nu} \sfD^{(2)}_\mu \Phi^\dagger \sfD^{(2)}_\nu \Phi-e^{-2\varphi_2}  M^2 \Phi^\dagger \Phi -\frac {\lambda}4 \left(\Phi^\dagger \Phi\right)^2\right]\label{Hscalar}
\ee
for a complex scalar field $\Phi$, where $ \sfD^{(2)}_\mu=\partial_\mu + \partial_\mu\varphi_2$. 

$S_{12}^{(c)}$ is confomal invariant. We can identify $\varphi_1$ with the $\varphi$ of the previous sections, while $\varphi_2$ is an additional dilaton field: with a suitable choice of the two `gauges' for $\varphi_1$ and $\varphi_2$, we can 
account for the two scales related to $\mathfrak c$ and $M^2$. But, beside the cosmological interpretation, the two dilaton fields  turn out handy in order to cancel both $S_C$ and $S_Q$, according to the previous scheme. In this context it is important to remember that setting $\varphi_1$ and $\varphi_2$ to constant values, the derivatives of the latter vanish and disappear in (\ref{Rscalar12},\ref{Smu}) and \eqref{Hscalar} (non-vanishing derivatives of the latter would be an obstacle for cancelation).

The second specification concerns the presence of counterterms in the one-loop effective actions. We have shown above how we can annihilate the quartic derivative terms corresponding to $S_C$ and $S_Q$ by subtracting counterterms that are necessary in order to cancel corresponding (trivial) trace anomalies. But in the previous section and also in Appendix C, while analyzing the $\mathfrak b$ cohomology, we have met several other (trivial) anomalies which needed countertems in order to be canceled. If these (trivial) anomalies are truly present in the theory we are obliged to modify the quantum action with the corresponding counterterms. The question that we would like to comment on here is the possibility that such counterterms contains quartic derivatives. If this were the case our efforts to cancel quartic derivatives in the action by means of one or two dilatons, as shown above, might be frustrated. The abstract analysis of the $\mathfrak b$ cohomology carried out in the previous section does not provide this information. To recover it we must delve into the flesh and bones of the theory. For instance, the counterterm \eqref{CL} which is necessary to cancel a trivial anomaly  due to the external $L$ field, is not worrying because it contains only one derivative. But the coboundaries generated by acting with $\delta_\omega$ on the  seven cochains of the list \eqref{list1}, as was explained in section 6.1, all have four derivatives. Fortunately they all contain derivatives of $\varphi$, therefore in a gauge in which $\varphi$ is constant the corresponding counterterms vanish\footnote{Even the last one of the list which gives rise to the cocycle numbered 16, see the end of section 6.1.}. Therefore in the framework discussed in this section they are irrelevant. There remain the three cocycles constructed with the three densities \eqref{list2}. To gather the relevant information we must descend from the abstract analysis of cohomology to the concrete task of working out explicitly  the trace anomalies of our theory, which we shall call down-to-earth approach.

\section{Weyl cocycles and Weyl coboundaries, the down-to-earth approach}

The form the Weyl cocycles (actually coboundaries) appear in the analysis of section \ref{s:weylcohomology} is not the familiar one we can meet in the literature. Weyl anomalies are usually a combination of quadratic curvature terms with precise coefficients, see below. In the literature for instance there is no mention of a Weyl anomaly whose density contains the square of the Ricci scalar alone. It is therefore necessary to understand this point and find a bridge between the traditional form of Weyl anomalies and the coboundaries found above, and between the abstract form of the $\mathfrak b$ cohomology and the familiar form of Weyl anomalies.

Let us start from the classical definition of the e.m. tensor for free matter fields interacting with a background metric, which is
\be
 T_{\mu\nu}= \frac 2{\sqrt g}\frac {\delta S}{\delta g^{\mu\nu}}\label{Tmunucl}
\ee
$S$ being the classical action. If the latter is Weyl-invariant the e.m. tensor is traceless. This follows from the classical Ward identity
\be
{\delta_\omega S}= \int d^4x\, \left(\frac {\delta S}{\delta g^{\mu\nu}} \delta_\omega g^{\mu\nu} +\sum_i \frac  {\delta S}{\delta f_i} \delta_\omega f_i\right)=0\label{completeWIcl}
\ee
where $f_i$ denotes generic matter fields. For infinitesimal $\omega$, $\delta_\omega g^{\mu\nu}= -2\omega g^{\mu\nu}, \delta_\omega f_i= -2 y_i \omega f_i$ (where $y_i$ is 0 for gauge fields, 1 for scalars and $\frac 32$ for fermions, etc.). If the matter fields are on shell (with the exception of the gauge fields), i.e.  $\frac  {\delta S}{\delta f_i} =0$, it follows that $T_{\mu\nu} g^{\mu\nu}=0$ due to the arbitrariness of $\omega$.

The problem we have to consider in the case of a classical action like $S^{(c)}$, defined by \eqref{totalactionc} where for simplicity of notation we drop the $\pm$ suffix,   is however more complicated because the metric is dynamical and there are multiple interaction terms that couple the fields in various ways. Invariance of the classical action $S^{(c)}$, under Weyl transformations is given by $\delta_\omega S^{(c)}=0 $. The procedure is the same as for \eqref{completeWIcl} except that the metric is not anymore a spectator, but we have to functionally differentiate also the gravity part of the action. The equation we obtained in \cite{BG24,bonora25} was 
\be
  R + 2\,{\mathfrak c}\,e^{-2\varphi}=2 \kappa\, e^{2\varphi} \,T^{(m)}, \quad\quad\quad{\rm where}\quad\quad T^{(m)}= g^{\mu\nu} T^{(m)}_{\mu\nu}\label{Riccieqmod}
\ee
which is the trace of the eom of $g_{\mu\nu}$, 
\be
R_{\mu\nu}-\frac 12 g_{\mu\nu}  \left(R -{\mathfrak c} e^{-2\varphi}\right)= 2 \kappa e^{2\varphi} T_{\mu\nu}^{(m)} \label{eom3}
\ee
Eq.\eqref{Riccieqmod} is the (the trace of the) eom of $g_{\mu\nu}$ with a source term represented by the trace of the e.m. tensor of the matter fields, $T_{\mu\nu}^{(m)}$, including the dilaton. However there are other possible interpretation of \eqref{Riccieqmod}.

First in the present theory we have to functionally differentiate, not only the EH part of the action, but $S^{(c)}_{gr}\equiv S^{(c)}_{EH}+S_C+S_Q$. This leads to
\be
\frac 2{\sqrt {  g}}\frac {\delta S^{(c)}_{gr}}{\delta {  g}^{\mu\nu}}   g^{\mu\nu} = T^{(m)}\label{Ricciquantum1}
\ee
where now $T^{(m)}$ is the e.m. of the matter fields, not including $\varphi$.  Since $S^{(c)}_{gr}$ can be expressed in terms of $\widetilde g_{\mu\nu}$ and tilded curvature and scalar tensors, we can also write
\be
\frac 2{\sqrt {  g}}\frac {\delta S^{(c)}_{gr}}{\delta {  g}^{\mu\nu}}  g^{\mu\nu}=\frac 2{\sqrt {\widetilde g}}\frac {\delta S^{(c)}_{gr}}{\delta {\widetilde g}^{\mu\nu}} \widetilde g^{\mu\nu} = \frac 1{2\kappa} \widetilde{\widetilde R}+2\frac {\mathfrak c}{\kappa} +\ldots \label{Ricciquantum2}
\ee
where the ellipses stand for quartic derivative terms that come from the variation of $S_C+S_Q$.
From the analysis of the kinetic term in section 5 it is evident that we cannot disentangle the dynamics of the metric from that of $\varphi$. Therefore we can interpret \eqref{Ricciquantum2} as the trace of the entangled equation of motion of the metric and $\varphi$ (the $\varphi$ eom does not provide additional information).

We understand \eqref{Ricciquantum2} as an equation that (partly) determines a background $g_{\mu\nu}$ and $\varphi$ solution over which we will carry out quantization. To be more precise: in \eqref{Riccieqmod} we consider only one possible non-trivial background field, the metric; in \eqref{Ricciquantum2} we may have two, the metric and the dilaton.   As for  the other matter fields we assume that they represent fluctuations about the null configuration. 
Therefore in general we might found our analysis on non-trivial background configurations for both the metric and the dilaton. But in this section, given its exploratory character, we focus on the simple case where the metric is the flat Minkowski one and the background of $\varphi$ is 0. Moreover  in order to parallel the quantization of section 5 we set $\mathfrak c_\pm=0$. Therefore at the classical background level eq.\eqref{Riccieqmod} reads: 0=0. The fluctuating part of $g_{\mu\nu}$, that is $h_{\mu\nu}$, as well as the fluctuating dilaton $\varphi$ in this approach will be spectator fields (in path integral jargon, we do not integrate over them). 

The last statement needs a more detailed explanation. 
First we split the action $S^{(c)}= S^{(c)}_0+ S^{(c)}_{int}$ into its free and the interacting part. $S^{(c)}_0$ contains only the kinetic quadratic terms of each separate species. $S^{(c)}_{int}$ contains all the interaction vertices (which are infinite in number because the metric and the dilaton have dimension 0). As shown in section 5, perturbative quantization is constructed with the propagators derived from $S^{(c)}_0$ and with the just-mentioned vertices, with the addition of gauge fixing and ghost actions. The e.m. tensor of the matter fields is obtained by expanding $S^{(c)}$ in $h_{\mu\nu}$ and selecting the first order in this expansion (the 0-th order is the action of the matter fields in the flat metric background). The first order splits into various pieces in which the matter fields are generally entangled, but, if we restrict ourselves to the lowest (non-interacting) order, the entanglement disappears and we find a sum of distinct e.m. tensors, one for each species of matter fields. For instance, the zero-th order e.m. tensor obtained in this way for fermions is
\be
 T^{(f)}_{\mu\nu}= \frac i4 \overline {\psi} \gamma_\mu {\stackrel{\leftrightarrow}{\partial_\nu}}\psi+(\mu\leftrightarrow\nu)- \eta_{\mu\nu}\frac i2  \overline \psi \gamma^\lambda {\stackrel{\leftrightarrow}{\partial_\lambda}}\psi, \label{Tmunufermion}
\ee
and for Abelian gauge fields is
\be
T^{(g)}_{\mu\nu} =-\frac 1{g^2}\left(F_{\mu\lambda} F_\nu{}^{\lambda} - \frac 14 \eta_{\mu\nu} F_{\lambda\rho}F^{\lambda\rho}\right)\label{Tmunugauge}
\ee
These e.m. tensors are conserved and traceless. For scalar fields the recovery of a divergenceless and traceless e.m. tensor is not so straightforward. This is because in $S^{(c)}$ conformal invariance of the scalar terms owes its occurrence to the interaction with the dilaton. However the corresponding interaction terms cannot appear at the level we are considering here (first order in $h_{\mu\nu}$). Therefore we have to resort to the same trick utilized in \cite{BG24}. 

For a complex scalar field $\Phi$ we expect a conserved and traceless expression like
\be 
T^{(s)}_{\mu\nu} =\partial_\mu \Phi^\dagger \,\partial_\nu \Phi+\partial_\nu \Phi^\dagger \,\partial_\mu \Phi- \eta_{\mu\nu}\, \partial_\lambda \Phi^\dagger \,\partial^\lambda \Phi+ \frac 13 \left(\eta_{\mu\nu} \square -\partial_\mu\partial_\nu\right) \Phi^\dagger \Phi\label{TmunuscalarPhi0th}
\ee
and an analogous formula for $H$. Such formulas are the improved canonical formulas. But, for the reason just explained, this is not what comes out spontaneously from the explicit expansion in $h_{\mu\nu}$.

Consider a free scalar field $\phi$ in dimension $\sfd$. Its improved (on-shell conserved and traceless) e.m. tensor is
\be 
T^{(s)}_{\mu\nu} =\partial_\mu \phi \,\partial_\nu \phi - \frac 12 \eta_{\mu\nu}\, \partial_\lambda \phi \,\partial^\lambda \phi+ \frac {\sfd -2}{4(\sfd -1)}  \left(\eta_{\mu\nu} \square -\partial_\mu\partial_\nu\right) \phi^2,\label{Tmunuscalarphi}
\ee
assuming the free equation of motion is $\square \phi=0$.
It can be derived with the improved canonical formula. But it can also be derived from the action
\be
S_\phi = \frac 12 \int d^\sfd x \,\sqrt{g} \left(\partial_\mu \phi \partial^\mu \phi +\frac {\sfd -2}{4(\sfd -1)} R\, \phi ^2\right) \label{scalaraction}
\ee
where $R$ is the Ricci scalar. This action is conformal-invariant. Applying \eqref{Tmunucl} to it, integrating by parts and setting $g_{\mu\nu} =\eta_{\mu\nu}$ one obtains \eqref{Tmunuscalarphi}. Now, using this, we do the following: to the free action of $\phi$ we add and subtract a term $\int d^4 x \,\sqrt{g}  R\, \phi ^2$ (with the appropriate coefficient). The expansion in $h_{\mu\nu}$ gives
\be
2  \int d^4 x \, h^{\mu\nu}  \left(\eta_{\mu\nu} \square -\partial_\mu\partial_\nu\right) \phi^2+ {\cal O}(h^2)\label{1stapproximant}
\ee
from which one can derive \eqref{Tmunuscalarphi}. Adding and subtracting the same term of course does not change anything, but while the lowest order approximant \eqref{1stapproximant} allows us to write \eqref{Tmunuscalarphi}, the term with opposite sign will remain and contribute an interaction term that was not present in the original action, and must be duly taken into account. All the terms of higher order in \eqref{1stapproximant} cancel one another. Said another way, we expand the metric on the RHS of \eqref{scalaraction}: the zeroth-order term is the free action of $\phi$ and is left unchanged by the addition of  $\int d^4 x \,\sqrt{g}  R\, \phi ^2$, the first order coefficient of  $h^{\mu\nu}$ identifies the on-shell conserved and traceless e.m. tensor \eqref{Tmunuscalarphi}; the first order term of  $-\int d^4 x \,\sqrt{g}  R\, \phi ^2$ is accounted for as an interaction term. All the higher order terms of   $\int d^4 x \,\sqrt{g}  R\, \phi ^2$ disappear. Of course to this new interaction term we have to add all the interaction terms of the scalar with the dilaton that were not detected by the expansion to the first order in $h_{\mu\nu}$. The scalar mass term must be also considered an interaction term.

Such acrobatics apply to any real or complex scalar field $\Phi$ or doublet $H$.  After that we have to consider also the gauge-fixing and ghost terms. As for the corresponding terms, related to the diffeomorphims and the conformal transformations, introduced in section 5, they concern $h_{\mu\nu}$ and $\varphi$, which are spectators, thus they will not play a role in the sequel. In the complete treatment of the $S^{(c)}$ theory we must also take account of the gauge fixing and ghost terms of the SM gauge symmetry. These are standard and we assume them implicitly when, below, we report the results for the relevant trace anomalies.  

After this long detour let us return to eq.\eqref{Ricciquantum1} and \eqref{Ricciquantum2} at the zeroth order level of the $h_{\mu\nu}$ expansion. The RHS of \eqref{Ricciquantum1} is the trace of a sum of  on-shell divergenceless and traceless e.m. tensors. This is because the various e.m. tensors at this level are disentangled from one another and the equations of motion without interactions reduce to the free ones for each separate species. Therefore the e.m. tensor for each species turns out to be traceless on shell. Thus the RHS of \eqref{Ricciquantum1} vanishes. The  LHS which is the RHS of \eqref{Ricciquantum2} with ${\mathfrak c}=0$ is given by:
\be
-\frac 3\kappa \square \varphi-\frac 6\gamma \square^2 \varphi+ \ldots\label{zerothorderh}
\ee
where the ellipses mean higher order terms in $\varphi$. This expression must vanish to. We assume that there are solutions of this higher derivative equations and they represent the value of the spectator $\varphi$.

The conclusion of this long digression is that all  we have to do is calculate trace anomalies of free fields coupled to various external potentials: Weyl fermions coupled to a metric, to gauge potentials and to scalars; gauge fields coupled to a metric; scalars or doublet of scalars coupled to a metric, to a gauge potential and to a dilaton. Thus we can utilize  the results one can find in the literature for free fields, \cite{christen78,christen79,parker78,duff1994,duff2020,parker2009,I}. As far as the metric dependence is concerned all the fields, fermions, gauge bosons, scalars, give rise to trace anomalies 
which are linear combinations of \eqref{quadweyl}, \eqref{gausbonnet} and \eqref{squareR}, with coefficients depending on the species and, for  \eqref{squareR}, also on the regularization method. 

As far as trace anomalies dependent on gauge fields are concerned we can have the following ones with density $\cal F$ given by
\be
T_e[V]=\tr\left( F_{\mu\nu}F^{\mu\nu}\right), \label{gaugeaction}
\ee
and
\be
T_o[V]=\varepsilon^{\mu\nu\lambda\rho}\tr \left(F_{\mu\nu}F_{\lambda\rho}\right).\label{chern}
\ee 
for a non-Abelian field $V_\mu$ with $F_{\mu\nu}= \partial_\mu V_\nu -\partial_\nu V_\mu+i [V_\mu,V_\nu]$, as well as others with Abelian gauge fields.  All such anomalies appear with a definite coefficient in front, depending on the field species which are integrated over (but not on the regularization used). Since in this section we are searching possible anomalies with four derivatives, these are of no interest here.

Fermions, via the Yukawa couplings, couple also to scalars and give rise to trace anomalies which are expressed in terms of scalar fields, see \cite{BG24}. Also these anomalies contain at most two derivatives, therefore they are not interesting for the issue considered in this section.

The previous ones as well as (\ref{quadweyl}, \ref{gausbonnet},\ref{pontryagin},\ref{squareR}),  are all anomalies that do not involve the dilaton field $\varphi$. But the $\mathfrak b -$cohomology analysis in our theory based on the action $S^{(c)}$, tells us that there are other possible trace anomalies which explicitly involve $\varphi$. For instance those with densities \footnote{ Here and below in eqs.(\ref{evenparitytracetilde}, \ref{R2}) repeated indices are saturated with the conformal invariant metric.}
\be
{\widetilde{\widetilde R}}^2\quad\quad {\rm and} \quad\quad   \widetilde  R_{\mu\nu} \widetilde R^{\mu\nu} 
\label{Rtildesquare}
\ee
or the square of the tilded Riemann tensor, are also consistent Weyl  cocycles.  But due to the presence of a dilaton field they are all trivial cocycles (or coboundaries).

The question that naturally arises is: what is the relation between these two different sets of cocycles? Let us take a concrete and simple enough case, the trace anomaly produced by a Dirac fermion in interaction with an external metric. The action
\be
S_D=\int d^4 x \, \left(\sqrt{g}\, i\overline {\psi}  
\gamma^a
 e_a^{\mu}
\left(\partial_\mu+\frac 12 \omega_\mu \right)\psi\right)( x)\label{Diracfermionaction}
\ee
 is Weyl invariant and, whether with perturbative or nonperturbative techniques, the result we find is the same:
\be
\ET(x) = \frac 1{16\pi^2} \left(\frac 1{72} R^2 -\frac 1{45} R_{\mu\nu}R^{\mu\nu} -\frac 7{360} R_{\mu\nu\lambda\rho}   R^{\mu\nu\lambda\rho}  \right)\label{evenparitytrace}
\ee
which is a combination of \eqref{quadweyl} and \eqref{gausbonnet}.

To see the appearance of cocycles like \eqref{Rtildesquare} by a calculations with Feynman diagrams is not simple because one has to take account also of the interaction between fermions and the dilaton. However there is a direct way, which consists in transforming the untilded into the corresponding tilded ones, i.e. in representing \eqref{evenparitytrace} as follows
\be
\ET(x) = \frac 1{16\pi^2} \left(\frac 1{72}\widetilde {\widetilde R}^2 -\frac 1{45}\widetilde R_{\mu\nu}\widetilde R^{\mu\nu} -\frac 7{360}\widetilde  R_{\mu\nu\lambda\rho}  \widetilde  R^{\mu\nu\lambda\rho}  +\ldots\right)\label{evenparitytracetilde}
\ee
where ellipses denote terms depending on derivatives of $\varphi$. For instance
\be
\sqrt{ g}\,R^2=\sqrt{\widetilde g}\left(\widetilde R^2 -12 \widetilde R \left(\widetilde \square \varphi + \partial\varphi \! \cdot\! \partial\varphi\right) +36\left(\widetilde \square \varphi\, \widetilde \square \varphi + 2  \widetilde \square \varphi \,  \partial\varphi \! \cdot\! \partial\varphi + \left( \partial\varphi \! \cdot\! \partial\varphi\right)^2\right)\right)\label{R2}
\ee
and similar, but much more cumbersome, expressions for the other square curvature terms. Let us recall that for constant $\varphi$ all the additional terms represented by the ellipses vanish.

It remains for us to consider the trace anomalies generated by a (complex) scalar field $\Phi$ or a scalar doublet $H$. The latter may interact also with gauge fields, giving rise to anomalies containing only two derivatives. Therefore let us consider as typical the 
case of a complex scalar field $\Phi$, whose relevant part of the action is
\be
S_\Phi^{(c)}&=& \int d^4 x \,\sqrt{g} \left[ g^{\mu\nu} \sfD_\mu \Phi^\dagger \sfD_\nu \Phi-e^{-2\varphi}  M^2 \Phi^\dagger \Phi \right]=\int d^4 x \,\sqrt{g}\,\Phi^\dagger \left( -\square- \square \varphi + \partial\varphi \! \cdot\! \partial\varphi\right)\Phi\0\\
&=& \int d^4 x \,\sqrt{g}\,\Phi^\dagger \left( -\square +\frac 16 (R-\widetilde R) \right)\Phi \label{Phiscalar}
\ee
where $ \sfD_\mu=\partial_\mu + \partial_\mu\varphi$. The free canonical e.m. tensor has been written down above, \eqref{TmunuscalarPhi0th}. But the perturbative approach would be rather cumbersome here.

A quicker way to obtain a response for the trace anomaly in this case is the non-perturbative method based on the heat kernel equation and the techniques introduced by Schwinger and DeWitt. In path integral language it is called integrating out the scalar field, i.e. viewing the one-loop path integral of $S^{(c)}_\Phi$ as the determinant of the Dirac operator (in fact the square root of the determinant of the squared Dirac operator). The Schwinger-DeWitt techniques allow us to extract such information as the trace of the e.m. tensor. We rely on this approach for a quick derivation of the result we are looking for. The trace of the em tensor one obtains is
\be
T_\Phi= \frac 1{16\pi^2} \left( \frac 1{180} \left( R_{\mu\nu\lambda\rho} R^{\mu\nu\lambda\rho}- R_{\mu\nu}R^{\mu\nu}  \right) +\frac 1{36} \widetilde R^2\right) \label{TPhi}
\ee
The anomaly ${\cal A}[\Phi]_\omega = \int d^4x \, \omega T_\Phi$ satisfies the consistency conditions separately for the sum of the first two terms and the last term. Therefore we can use separate counterterms for each of them. In particular to cancel the last terms we can use
\be
 {\cal C}_\Phi =- \frac 1 {576 \pi^2} \int d^4x \sqrt {\widetilde g} \, \varphi \,{ \widetilde{\widetilde R}}^2\label{CPhi}
 \ee
 Here $\varphi$ represent a generic dilaton field (it could represent $\varphi_2$  introduced in the previous section). But ${\cal C}_\Phi$ is precisely the conterterm $C(Q)$ of \eqref{Q2counter}, needed to cancel the quartic derivatives present in $S_Q$ by fixing $\varphi$ at a suitable constant value. 
 
 On the other hand we have seen above that an analogous cancelation of the quartic derivatives introduced by $S_C$ can be obtained in a similar way by means of the counterterm $C(W)$, \eqref{quadweylcounter}.

As for the (trivial) anomaly whose density is $\widetilde\square \widetilde R$  or $\square R$, depending on the approach, when the dilaton field in the counterterm for the first, or in the WZ term for the second, is set to a constant value, they both vanish for the integrand reduces to a total derivative. 

A final remark concerns the number of different dilatons needed in the previous construction. We need two different dilatons for the conterterms $C(W)$ and ${\cal C}_\Phi$ and an additional dilaton in order to cancel the anomaly whose density is $\sim \left( R_{\mu\nu\lambda\rho} R^{\mu\nu\lambda\rho}- R_{\mu\nu}R^{\mu\nu} +\square R \right) $ coming from \eqref{TPhi}, by means of an appropriate WZ term.

Summarizing the question of quartic derivative terms in the action, the situation is as follows. The aim of this section was to be able to show that the quartic derivative terms $S_C$ and $S_Q$, which are required by renormalization in order to keep the divergent Feynman diagrams under control, do not carry  physical ghosts inside the theory. To this end we have embedded them in a conformal invariant theory: our point is to be able to show that in a specific gauge the quartic derivatives disappear. This may be done thanks to the presence of trace anomalies at one-loop in the theory. Being trivial these anomalies can be canceled by counterterms that restore one-loop conformal invariance. It so happens that the counterterms can be chosen in such a way that a specific choice of conformal gauge for the dilaton field(s) cancel exactly both  $S_C$ and $S_Q$. This can be done both for the approach compliant with the algebraic renormalization based on the BPHZL regularization scheme, as well as for the approach we have called down-to-earth. 

{\bf Comment.} A comment is in order concerning odd-parity trace anomalies, such as \eqref{pontryagin} and \eqref{chern}, to avoid misunderstandings. These anomalies could be canceled by the corresponding WZ term. However these anomalies have imaginary coefficients, which implies that the em tensor picks up a non-hermitean part due to quantization and a consequent violation of unitarity. Of course such anomalies could be canceled by means of the corresponding WZ term. However, on the basis of the previous discussion, this would require the addition of imaginary terms to the effective action, which would compromise  its hermiticity. In short anomaly cancelation by means of WZ terms is not viable in this case. On the other hand it was clarified in \cite{BG24} that such anomalies are the symptoms of a disease, not the disease itself. The latter is represented by the cohomological obstruction to the existence of Weyl fermion propagators due to the presence of the Pontryagin and Chern class in the index of the Dirac-Weyl operator, as it is made explicit by the family's index theorem of Atiyah and Singer. There is no way to bypass this obstacle but with a mirror construction like the one presented in this paper in section 2.

 \section{Conclusion}
 
The theory proposed in this paper incorporates both SM and gravity in a form that avoids all the type-O anomalies, as was first shown in \cite{BG24}. Following \cite{bonora25} it has been presented here in a simplified  version involving only one metric, instead of two. It preserves nevertheless the basic structure in two sectors, left and right, with mirror fermions and scalars, each sector with its own $SU(3)$ and $U(1)$ gauge fields, while the $SU(2)$ gauge fields as well as the metric are in common. In the first part of the paper some basic ideas, already presented in \cite{bonora25}, have been reviewed: in particular the interpretation of the right sector as the dark matter one, a rather attractive and reasonable idea, and the proposal of Weyl symmetry as a fundamental symmetry and its possible connections with cosmology as well as its consequences on the theoretical side that may affect unitarity and renormalization. The present paper has focused mostly on the quantizaion of the model. The algebraic renormalization based on the BPHZL subtraction scheme has been developed to a considerable extent: the propagators and vertices have been analyzed and shown that they essentially satisfy the convergence properties required by the BPHZL method; the ST identity for diffeomorphisms+ Weyl transformations has been written down, together with the gauge fixing and ghost equations; a formal proof of gauge independence has been given; the nilpotent linearized ST operator has been defined and the corresponding cohomology algebra completely determined. It has been shown that in this theory conformal anomalies are trivialized by counterterms. It has been also stressed that such counterterms cannot be swept under the carpet, they actually can play a major role in canceling  quartic derivative terms in the action, which may be the origin of the presence in the theory of non-physical particles. In any case this is only a necessary condition to prove unitarity of the theory, but nevertheless a significant indication that it is not  impossible to do it.

Besides, there are still many unexplored aspects. The first is the completion of the renormalization process started here, for the complete theory including all the matter fields. What is missing in particular is the definition of the renormalization conditions, the Callan-Symanzik and renormalization group equations, an uncontroversial proof of unitarity not only at the first level of approximation but at all loop orders (the present paper is often limited to one-loop considerations). Another basic progress would be to extend these results to the same theory quantized over a nontrivial vacuum, such as the one of \cite{bonora25}. 

\vskip 1cm

{\bf Note added.} One of the reviewers of this paper has prompted a comment relative to the following work:  N.C. Tsamis and R.P. Woodard, {\it No New Physics in Conformal Scalar-Metric Theory}, Annals Phys. {\bf 168} (1986) 457, which claims that invariantly regulated conformal scalar-metric theories are physically equivalent
to their pure metric analogs. It illustrates a point of view somewhat opposite to the one expressed in the present paper, but a conformal scalar-tensor theory (an example of which is given in eq.(160))  is not the theory considered in the present paper, which is essentially based on the role of dilatons. The scalar of the scalar-tensor theory is a dimensionful field (of dimension 1 in 4d), while a dilaton is dimensionless. Although there is some overlap between the two papers, for instance the absence of trace anomalies, they are quite different. The crucial point, highlighted in the present paper, for which a conformal theory of matter and gravity is preferable and maybe even compulsory, is explained in section 4.1. The dilaton has a physical role, it is a bridge between different scales, without which it would be impossible to quantize the theory due to the gigantic difference between the quantization scale of the SM and that of gravity. It is only by means of one or more dilatons that we can bring the two scales at  the same level, where quantizing SM and gravity together makes sense.

\section{Appendix A. Notations} 

First we introduce the basic two-index projectors
\be
\omega_{\mu\nu} =\frac {p_\mu p_\nu}{p^2}, \quad\quad \theta_{\mu\nu}=\eta_{\mu\nu} - \omega_{\mu\nu}\label{omegatheta}
\ee
Then in the representation space of two-index symmetric tensors we introduce the projectors
\be
&&P^{(2)}_{\mu\nu\lambda\rho} = \frac 12 \left(\theta_{\mu\lambda} \theta_{\nu\rho} + \theta_{\mu\rho}\theta_{\nu\lambda} \right) - \frac 13 \theta_{\mu\nu} \theta_{\lambda\rho} \0\\
&&P^{(1)}_{\mu\nu\lambda\rho} =\frac 12 \left( \theta_{\mu\lambda} \omega_{\nu\rho}+ \theta_{\mu\rho} \omega_{\nu\lambda} + \theta_{\nu\lambda} \omega_{\mu\rho}+ \theta_{\nu\rho}\omega_{\mu\lambda}\right)\0\\
&& P^{(0)}_{\mu\nu\lambda\rho} = \frac 13 \theta_{\mu\nu} \theta_{\lambda\rho} \0\\
&&\overline P^{( 0)}_{\mu\nu\lambda\rho} =\omega_{\mu\nu}\,\omega_{\lambda\rho}\label{2projectors}
\ee
It is easily verified that they are orthogonal projectors: any two of them satisfy
\be
P_i^{\mu\nu}{}_{\lambda\rho} P_j^{\lambda\rho\sigma\tau} = \delta_{ij}P_i^{\mu\nu\sigma\tau}  
 \ee
 and
\be
\left( P^{(2)} + P^{(1)}+P^{(0)}+\overline P^{(0)}\right)_{\mu\nu\lambda\rho}=\frac  12 \left( \eta_{\mu\lambda}\eta_{\nu\rho}+\eta_{\mu\rho} \eta_{\nu\lambda}\right)\label{identity}
\ee
which is the identity in the space of two-index symmetric tensors. To guarantee completeness we need also another operator
\be
T^{(0)} =\frac 1{\sqrt 3}\left(\theta_{\mu\nu} \omega_{\lambda\rho} + \theta_{\lambda\rho} \omega_{\mu\nu}\right)\label{T0}
\ee
It is customary in the literature to consider separately the two terms on the RHS of this equation, but in this paper we need only their sum. $T^{(0)}$ is not a projector. It is orthogonal to $P^{(2)}$ and $ P^{(1)}$ but not to $P^{(0)}$ and $\overline P^{(0)}$; rather
\be
T^{(0)} T^{(0)} = P^{(0)} +\overline P^{(0)}\label{T0square}
\ee

\section{Appendix B. Non-renormalization of gauge  fixing and ghost equations}

The aim of this appendix is to study the renormalization of the gauge fixing conditions and the ghost equations. Let us start from \eqref{gfcond1}, which is certainly true when $\Gamma=S_{tot}$, see \eqref{Stot}. The problem is to see whether and how this equation is `deformed' by quantization:
\be
\frac{\delta \Gamma}{\delta b^\mu} =\frac{\delta S_{tot}}{\delta b^\mu}+ \sum_{n=1}^\infty \hslash^n \Delta_\mu^{(n)}\label{deltaGammabquantum}
\ee
$\Delta^{(1)}$ has dimension 2 and ghost number 0. The quantum action principle tells us that it is a polynomial of the fields, thus it may contain linearly $b_\alpha$ plus a  combination of fields  $\phi_i$,  $F_\mu(\phi_i)$, with the exclusion of $b_\alpha$ and $b$, with dimension 2 (possibly including $\kappa$) and ghost number 0 (for instance $ \frac 1{\sqrt{\kappa}} \partial_\mu \varphi$), that is
\be
\Delta_\mu^{(1)}= F_\mu(\phi_i) +c_{\mu\alpha} b^\alpha\label{Deltamu1}
\ee
where $c_{\mu\alpha}$  are constants. Since $\frac {\delta}{\delta b_{\alpha}}(x) \frac {\delta}{\delta b_{\mu}}(y)=  \frac {\delta}{\delta b_{\mu}}(y)\frac {\delta}{\delta b_{\alpha}}(x)$, it follows that
$c_{\mu\alpha}= c_{\alpha\mu}$, then we can write
\be
\Delta_\mu^{(1)}=  \frac{\delta}{\delta b^\mu}\int d^4 x \left( b^\mu  F_\mu(\phi) +\frac 12 c_{\mu\nu}b^\mu b^\nu
\right)=\frac{\delta}{\delta b^\mu} C( b_\alpha, \phi_i,\kappa)\label{Deltamu2}
\ee
By subtracting the counterterm $C( b_\alpha, \phi_i)$ we restore the original form of eq.\eqref{gfcond1} at one-loop. Next, supposing that this has been done to order $n-1$, we can repeat the previous argument to order $n$ and prove the validity of \eqref{gfcond1} by induction.

As for \eqref{gfcond2} the procedure is the same. The only difference is that the counterterm is
\be
C(b,\phi_i)= b F(\phi_i, b) +\frac 12 b^2\label{counterterm2}
\ee
where the set of $\phi_i$ excludes $b_\alpha$ and $b$. 

Concerning \eqref{ghosteq1} one could repeat the same things as for \eqref{gfcond1}, except that the counterterm can only be linear in $\overline \xi^\alpha$ because there cannot be any quadratic term in the same field. Same arguments for  \eqref{ghosteq2}.

\section{Appendix C. Cohomology of external fields}

The aim of this Appendix is to analyse the $\mathfrak b$-cohomology involving external fields. The $\mathfrak b$ transformation of the external fields are given in \eqref{btransf}

The action of ${\mathbb S}$ on $\Gamma$ produces terms  of dimension 4 and ghost number 1, see \eqref{Deltan}. $\Delta^{(n)}$ are integrated polynomials of the fields and their derivatives. In this Appendix we are interested in possible terms that contain the external fields. Let us consider the external field $K$, which has dimension 4 and ghost number -1. Let us denote by $\Delta(K)$ the part of $\Delta^{(n)}$ linear in $K$ \footnote{In the sequel, to simplify the notation, we write $\partial\xi$ for $\partial_\lambda \xi^\lambda$, and $A\!\cdot\! B$ for $A^\lambda B_\lambda$.}. Let us start from a particular form for it
\be
 \Delta(K)= \int \left( a\, \partial \xi \,\omega\, K + b\, \xi\!\cdot\! \partial \omega\, K\right)\label{DeltaK}
 \ee 
 where $a,b$ are constants numbers. Consistency requires that applying $\mathfrak b$ we  get 0:
 \be
 0={\mathfrak b} \Delta(K)=\! \int\! \left( a\left( \xi^\lambda \partial_\lambda\partial_\alpha \xi^\alpha\,\omega\, K - 2 \partial\xi \,\xi\!\cdot\!\partial \omega \, K \right) + b \,\xi\!\cdot\! \partial\xi\!\cdot\!\partial\omega\, K+ {\rm terms \,\, independent\,\, of \,\, K}\right)\label{bDeltaK}
 \ee 
Since the polynomials proportional to $a$ and $b$ are independent we must conclude that the only allowed solution is $a=b=0$. But this analysis is not complete because we must consider the case where $a$ and $b$ are functions depending on $\varphi$, a field which is dimensionless and ghost number 0. In this case however the expression \eqref{DeltaK} is insufficient and the most general form is
\be
 \Delta'(K)= \int \left( a(\varphi)\, \partial \xi \,\omega\, K + b(\varphi)\, \xi\!\cdot\! \partial \omega\, K+ \frac{\delta c(\varphi)}{\delta \varphi} \xi\!\cdot\! \partial\omega \, K\right)\label{Delta'Kc}
 \ee 
 where $c=c(\varphi)$ is another independent function of $\varphi$. The action of $\mathfrak b$ has the following effect
 \be
  0&=&{\mathfrak b} \Delta(K)=\! \int\! \left( a\left( \xi^\lambda \partial_\lambda\partial_\alpha \xi^\alpha\,\omega\, K - 2 \partial\xi \,\xi\!\cdot\! \partial \omega \, K \right) + b \,\xi\!\cdot\! \partial \xi \!\cdot\! \partial \omega\, K\right.\0\\
  &&  \left.+ 2\frac{\delta a}{\delta \varphi} \xi\!\cdot\! \partial \varphi\, \partial\xi \, \omega\, K + \frac{\delta b}{\delta \varphi}\left( 2 \xi\!\cdot\!  \partial \varphi\, \xi\!\cdot\! \partial \omega  + \omega \xi\!\cdot\!  \partial \omega \right) K -2 \frac{\delta c}{\delta \varphi}\xi \!\cdot\!  \partial \xi \!\cdot\! \partial \omega\, K\right. \0\\
  &&+ \left. \frac{\delta c}{\delta \varphi} \xi \!\cdot\!  \partial  \xi\!\cdot\! \partial \varphi\,\omega \, K+{\rm terms \,\, independent\,\, of \,\, K}\right)\label{bDelta'Kc}
 \ee 
The only solution is $a=b=\frac{\delta c}{\delta \varphi}=0$. 

From this analysis it follows that the quantum corrections $\Delta^{(n)}$ cannot depend on the field $K$. From this example, which is relatively simple, we learn a lesson: the most complete treatment based on the expression \eqref{Delta'Kc} does not change the conclusions on $a$ and $b$ based on the expression \eqref{DeltaK}. If, like in this case, the conclusion is $a=b=0$, this remains true also for the most general case. In other words, to conclude the most general analysis one has to consider only the third term of \eqref{Delta'Kc}.

Let us consider next the case of $K_{\mu\nu}$, which has dimension 4 and ghost number $ -1$. The most general expression with constant coefficients is
\be
\Delta (K_{\mu\nu}) =  \int \left( a\,(\partial_\mu\xi_\nu+\partial_\nu\xi_\mu)  K^{\mu\nu}\,\omega\right) +b \left((\partial_\mu\omega \xi_\nu + \partial_\nu\omega \xi_\mu) K^{\mu\nu}\right) \label{DeltaKmunu}
\ee
The equation one gets from ${\mathfrak b}\Delta(K_{\mu\nu})=0$ contains a number of monomials proportional to $a$ and a number of monomials proportional to $b$. These two sets do not have 
any term in common. Thus $a=b=0$, trivially. It remains for us to consider the term
\be
\Delta (K_{\mu\nu} ) =\int  \frac{\delta c}{\delta \varphi}\left(\partial_\mu \varphi\,\omega\, \xi_\nu+ \partial_\nu \varphi\,\omega\, \xi_\mu\right) K^{\mu\nu}\label{DeltaKmunu}
\ee
It is easy to see that this is not an invariant for any $\frac{\delta c}{\delta \varphi}$. Thus $K_{\mu\nu}$ cannot appear in the quantum corrections $\Delta^{(n)}$. 

The case of the external field $L$ is less trivial. $L$ has dimension 4 and ghost number -2. Let us start with constant coefficients
\be
\Delta(L) = \int \left( a\, \partial\xi \, \xi\!\cdot\! \partial\omega \, L +b \, \xi\!\cdot\! \partial\xi \!\cdot\! \partial \omega\, L\right)\label{DeltaL}
\ee
Acting with $\mathfrak b$ we get
\be
0={\mathfrak b}\, \Delta (L) = a\, \int  \partial\xi\, \xi\!\cdot\! \partial\xi \!\cdot\! \omega \, L\label{bDeltaL}
\ee
Thus $a=0$ and the second term on the RHS of \eqref{DeltaL} satisfies the consistency conditions with generic constant $b$. This term is admitted in the quantum corrections $\Delta^{(n)}$. However we have at our disposal the field $\varphi$ and it is simple seeing that the counterterm 
\be
C(L)=\int \xi\!\cdot\! \partial\xi \!\cdot\! \varphi\,L, \quad\quad {\rm yields} \quad\quad {\mathfrak b}\, C(L)= \int  \partial\xi\, \xi\!\cdot\! \partial\xi \!\cdot\! \omega \, L\label{CL}  
\ee
Therefore $\Delta(L) $ with $a=0$ and $b$ constant is a coboundary. 

It remains for us to show that the dependence of the coefficients $a$ and $b$ on $\varphi$ and the addition of 
\be
\Delta'(L) = \int  \frac{\delta c}{\delta \varphi} \xi \!\cdot\!  \partial \varphi\,  \xi \!\cdot\!  \partial\omega \, L\label{Delta'L}
\ee
does not change the previous conclusion. In particular we must check that the following equation is verified:
\be
0= {\mathfrak b} \Delta'(L) = \int  \left(\frac{\delta^2 c}{\delta \varphi^2}\,\omega \, \xi \!\cdot\!  \partial \varphi\, \xi \!\cdot\!  \partial\omega , L - \frac{\delta c}{\delta \varphi}\,\xi\!\cdot\! \partial \xi \!\cdot\!  \partial \varphi\,\xi \!\cdot\!  \partial\omega \, L+ \frac{\delta c}{\delta \varphi}\,\xi \!\cdot\!  \partial \varphi\,  \xi\!\cdot\! \partial \xi \!\cdot\!  \partial \omega\, L\right)\label{bDelta'L}
\ee
It requires that $\frac{\delta c}{\delta \varphi}=0$. Eq. ${\mathfrak b} \Delta(L)=0$ with $a$ and $b$ being $\varphi$-dependent functions, 
yields $\frac{\delta a}{\delta \varphi}=\frac{\delta b}{\delta \varphi}=0$. This confirms that $\Delta(L) $ with $a=0$ and $b$ constant, as well as $\frac{\delta c}{\delta \varphi}=0$, is a cocycle. In fact, as we have seen, it is a coboundary.

The remaining external field is $L_\alpha$, which dimension 5 and ghost number $-2$. The candidate cycles with constant coefficients is
\be
\Delta(L_\alpha) = \int \left( a\, \partial \xi\, \omega\, \xi^\alpha L_\alpha+ b\, \xi\!\cdot \!\partial \omega\,\xi^\alpha L_\alpha+ c\, \omega\, \xi\!\cdot\! \partial \xi^\alpha l_\alpha\right). \label{DeltaLalpha}
\ee 
The relevant consistency equation is
\be
0&=&{\mathfrak b}\, \Delta(L_\alpha)=\int \left( a\, \left( 2\xi^\alpha\partial_\alpha \partial_\beta \xi^\beta \, \omega\, \xi^\lambda L_\lambda-2 \partial\xi \, \xi \!\cdot\!\partial \omega\,  \xi^\lambda L_\lambda+ \partial \xi \, \xi\!\cdot\! \partial \xi^\lambda l_\lambda\right)\right.\0\\
&&\left. +b\,\left( \xi\!\cdot\!\partial\xi \!\cdot\! \partial\omega \,\xi^\lambda L_\lambda+ \xi\!\cdot\!\partial\omega\, \xi\!\cdot\!\partial\xi^\lambda L_\lambda\right)+c\,\left(2 \xi\!\cdot\!\partial\omega\, \xi\!\cdot\!\partial\xi^\lambda L_\lambda-2 \,\omega\,\xi\!\cdot\!\partial\xi \!\cdot\! \partial\xi^\lambda L_\lambda \right)\right)\label{bDelatLalpha}
\ee
which implies $a=b=c=0$. Finally we can consider the case  in which $a,b,c$ are $\varphi$-dependent functions and add for completeness the term
\be
\Delta'(L_\alpha) = \int \frac{ \delta d}{\delta\varphi} \, \xi\!\cdot\!\partial\varphi \, \omega\,  \xi^\lambda L_\lambda\label{Delta'Lalpha}
\ee
but the conclusion does not change: $a=b=c=\frac{ \delta d}{\delta\varphi}=0$. Thus $L_\alpha$ cannot appear in the quantum corrections $\Delta^{(n)}$. 

In conclusion, the cohomology of $\mathfrak b$, as far as the external fields are concerned, is trivial. There is only one coboundary, given by the RHS of eq.\eqref{CL}, which can be reabsorbed by subtracting the counterterm $C(L)$.

It must be remarked that the subtraction of the counterterm $C(L)$ does not affect neither the gauge fixing conditions (\ref{gfdiffconf}, \ref{gfconf}) nor the ghost equations (\ref{ghosteqfin1},\ref{ghosteqfin2}).

\end{document}